\documentclass[utf8]{frontiersFPHY} 

\setcitestyle{square} 
\usepackage{url,hyperref,lineno,microtype,subcaption}
\usepackage[onehalfspacing]{setspace}
\usepackage{ulem}
\newcommand{\kep}{{\it Kepler}}

\def\keyFont{\fontsize{8}{11}\helveticabold }
\def\firstAuthorLast{Mathur {et~al.}} 
\def\Authors{Savita Mathur\,$^{1,2,*}$, Rafael A. Garc\'ia\,$^{3,4}$, Lisa Bugnet\,$^{3,4}$, \^{A}ngela R. G. Santos$^{5}$, Netsha Santiago$^{6}$, and Paul G. Beck$^{1,2,7}$}

\def\Address{$^{1}$Instituto de Astrof\'{\i}sica de Canarias, La Laguna, Tenerife, Spain \\
$^{2}$Dpto. de Astrof\'{\i}sica, Universidad de La Laguna, La Laguna, Tenerife, Spain \\
$^{3}$IRFU, CEA, Universit\'e Paris-Saclay, Gif-sur-Yvette, France\\
$^{4}$AIM, CEA, CNRS, Universit\'e Paris-Saclay, Universit\'e Paris Diderot, Sorbonne Paris Cit\'e, F-91191 Gif-sur-Yvette, France\\
$^{5}$Space Science Institute, 4750 Walnut Street, Suite 205, Boulder CO 80301, USA\\
$^{6}$ University of Puerto Rico at Cayey, Puerto Rico\\
$^{7}$Department for Geophysics, Meteorology and Astrophysics, Institute of Physics, Karl-Franzens University of Graz, Universit\"atsplatz 5/II, 8010 Graz, Austria}

\def\corrAuthor{Savita Mathur}

\def\corrEmail{smathur@iac.es}

\begin{document}
\onecolumn
\firstpage{1}

\title[Magnetic activity and oscillation]{Revisiting the impact of stellar magnetic activity on the { detectability} of solar-like oscillations by {\it Kepler}} 

\author[\firstAuthorLast ]{\Authors} 
\address{} 
\correspondance{} 

\extraAuth{}

\maketitle

\begin{abstract}



Over 2,000 stars were observed for one month with a high enough cadence in order to look for acoustic modes during the survey phase of the \kep\ mission. Solar-like oscillations have been detected in { about} 540 stars. { The question of why no oscillations were detected in the remaining stars is still open.} Previous works explained the non-detection of modes with the high level of magnetic activity of the stars. However, the sample of stars studied contained some classical pulsators and red giants that could have biased the results.
In this work, we revisit this analysis on a cleaner sample of main-sequence solar-like stars that consists of { 1,014} stars.  First we compute the predicted amplitude of the modes of that sample and for the stars with detected oscillation and compare it to the noise at high frequency in the power spectrum. We find that the stars with detected modes have {  an amplitude to noise ratio larger than 0.94.} { We measure} reliable rotation periods and the associated photometric magnetic index for 684 stars { out of the full sample and in particular for 323 stars where the amplitude of the modes is predicted to be high enough to be detected}. 
We find that { among these 323 stars 32\% of them} have a level of magnetic activity larger than the Sun during its maximum activity, explaining the non-detection of acoustic modes. Interestingly, magnetic activity cannot be the primary reason responsible for the absence of detectable modes in the remaining { 68$\%$} of the stars { without acoustic modes detected and with reliable rotation periods}. Thus, { we investigate}  metallicity, inclination angle of the rotation axis, and binarity as possible causes of low mode amplitudes.
Using spectroscopic observations for a subsample, we find that a low metallicity could be the reason for suppressed modes. No clear correlation with binarity nor inclination { is found}. { We also derive the lower limit for our photometric activity index (of 20-30\,ppm) below which rotation and magnetic activity are not detected.} Finally, with our analysis we conclude that stars with a photometric activity index larger than 2,000\,ppm have 98.3\% probability of not having oscillations detected.

%


\tiny
 \keyFont{ \section{Keywords:} Asteroseismology, Stellar Rotation, Magnetic Activity, Main-sequence stars, Solar-like oscillations} 
\end{abstract}

\section{Introduction}

Solar-type stars show acoustic oscillations that are intrinsically damped and stochastically excited by near-surface convection \citep[e.g.][]{1977ApJ...212..243G}, which is also an important ingredient for magnetic activity { \citep[][references therein]{2017LRSP...14....4B}}.

In the Sun, the properties of the acoustic modes are sensitive to the varying magnetic activity \citep[e.g.][]{1985Natur.318..449W,1990Natur.345..322E,2015MNRAS.454.4120H}. In particular, as a result of magnetic structures in the solar photosphere being strong absorbers of the acoustic waves \citep[e.g.][]{1987ApJ...319L..27B,2002A&A...387.1092J}, the amplitudes of the acoustic modes decrease with increasing activity level { \citep[e.g.][]{2000MNRAS.313...32C,2000ApJ...531.1094K,2015MNRAS.454.4120H,2018SoPh..293..151K}}. { Already observed in the Sun,} such activity-related variations were first discovered in a star other than the Sun, HD~49933 observed by the CoRoT satellite, by \citet{2010Sci...329.1032G}. Searching for such signatures of magnetic activity in {\it Kepler} data, it was found that other solar-type stars, in particular KIC~8006161 and KIC~5184732, show evidence for a decrease in the mode heights with increasing activity \citep{2017A&A...598A..77K,2018A&A...611A..84S,2018ApJS..237...17S}. 

The suppression of the acoustic modes by magnetic activity hampers their detection and, thus, magnetic activity may have important impact on the detectability of solar-type pulsators. In order to test this, \citet{2011ApJ...732L...5C} analyzed the first month of short-cadence {\it Kepler} data of about 2,000 stars. The authors successfully detected solar-type oscillations in only $\sim$\,540 stars and found that as the activity level increases the number of detections decreases significantly. Thus, the non-detections were attributed to high activity levels. However, they were limited to one month of data, which on the one hand hampers the detection of acoustic oscillations and on the other hand, may bias the activity level estimates. Also, the sample of about 2,000 supposedly solar-like stars was polluted by other types of stars, in particular classical pulsators and red giants. Furthermore, since then, more accurate photometric proxies of magnetic activity have been developed. Indeed in the past several indexes were used mostly based on the standard deviation of the time series allowing to measure the variability of the stars. But this variability can be due to different phenomena such as oscillations, convection, or magnetic activity \citep[e.g.][]{2010Sci...329.1032G,2010ApJ...713L.155B,2011AJ....141...20B,2011ApJ...732L...5C}. When measuring the magnetic activity index, we need to make sure that the variability is related to magnetic activity and not to other phenomena.

In this work, we analyze the latest data release of 2,576 targets observed in short-cadence by {\it Kepler}. Section~\ref{sec:obs} describes the data used and the selection of the sample. 
In Section~\ref{sec:data_an}, we describe the methods used to measure the surface rotation periods and the magnetic activity proxy. The surface rotation period is obtained by implementing the analysis in \citet{2014A&A...572A..34G} and \citet{2016MNRAS.456..119C}, which combines a time-frequency analysis with the auto-correlation function. Knowing the rotation period $P_\text{rot}$, we then measure the photometric magnetic activity index, which corresponds to the average value of the standard deviation of subseries of length $5\times P_\text{rot}$ \citep[$S_{\rm ph}$; e.g.][]{2014A&A...562A.124M}. Section~\ref{sec:results} provides the results of this analysis. In Section~\ref{sec:discussion}, we study the impact of magnetic activity, metallicity, and stellar inclination on the detection of the acoustic modes, searching for correlations between those properties. Finally, Section~\ref{sec:conclusions} summarizes the main conclusions.

\section{Observations: Defining the sample}\label{sec:obs}

\subsection{Data calibration}
{ In this work,} the {\it Kepler} light curves analyzed for rotation were generated from the { long-cadence (29.4244 min) pixel} files available at the Mikulski Archive for Space Telescopes (MAST) website. The integration of the stellar signal is done in masks larger than the usual ones used for exoplanet research in order to reduce the drifts induced by the displacement of the targets on the {\it Kepler} CCD. To do so, we take all the pixels centered at the coordinates of the star up to a radius for which a reference value per pixel, calculated as the 99.9 percentile of the weighted flux, is greater than 100. Then to correct for the remaining drifts, jumps and to stitch all the quarters together we follow the \kep\ Asteroseismic Data Analysis and Calibration Software (KADACS)\footnote{{ time-series are available at MAST via https://doi.org/10.17909/t9-cfke-ps60.}} methods described in \citet{2011MNRAS.414L...6G}. The lightcurves produced in this way are low-pass filtered using three different cut-off periods at 20, 55, and 80 days. The filters are made in such a way that the transfer function is one up to the mentioned period and then it smoothly decrease to zero at the double period, ensuring some transmission of the signals until 40, 110 and 160 days respectively. Finally, gaps up to 20 days are interpolated using a multi-scale cosinus transform using inpainting techniques as described in \citet{2014A&A...568A..10G} and \citet{2015A&A...574A..18P}.
As a comparison, we also use standard Pre-search Data Conditioning multi scale Maximum A-Posteriori light curves \citep[PDC-msMAP, e.g.,][]{2012PASP..124..985S,2012PASP..124.1000S,ThompsonRel21} to look for rotation periods and we compare the results with the three previously explained light curves. As shown in \citet{2013ASPC..479..129G}, it is more reliable to use different light curves prepared using completely different ways because each calibration system can fail for different reasons in particular targets. { For the analysis of the rotation, we used all the quarters available in long cadence, i.e. 4 years of continuous data, except those quarters where the star falls on one of the bad modules of \kep.  }

\subsection{Target selection}


The original sample was selected based on the target list of the Working Group 1 { (WG1)} of the {\it Kepler} Asteroseismic Science Consortium (KASC\footnote{{ http://kasoc.phys.au.dk/}}) that focuses on solar-like stars. { During the first 10 months of the mission, a survey phase was led in order to help the selection of the most promising targets for planet search or for asteroseismic analyses.} As there were only 512 slots available to download the short-cadence data (due to the { limited} downlink bandwith),  each month a different sample of stars was targeted. Thus we start with a sample of 2,576 targets for which data was collected { with a 58.85\,s} cadence. 

First we remove four stars that do not have any effective temperature values. By studying the power spectrum, we can see that a fraction of these stars are classical pulsators or red giants. { Classical pulsators usually present very high amplitude peaks in the power spectrum. We automatically discard stars which power spectra have peaks with amplitudes larger than  $10^7 {\rm ppm}^2 \mu {\rm Hz}^{-1}$.} As  some of the remaining stars have an effective temperatures larger than 7,500K, we visually check them and find that they indeed have classical-pulsators like peaks. Since the amplitudes of these peaks are below our threshold  they were not flagged by our automatic search. 

The next step is to cross-check our sample with the legacy catalog of the {\it Kepler} red giants (Garc\'{i}a et al. in prep.). This catalog makes use of known red giants from the literature \citep[e.g.][]{2013ApJ...765L..41S,2016ApJ...827...50M,2018ApJS..236...42Y}. In particular it also uses new techniques to classify the stars such as the FliPer method \citep{2018A&A...620A..38B} or neural networks \citep{2018MNRAS.476.3233H} as well as a visual inspection of all the stars. This leads to a sample of about 1,900 stars that should most probably be on the main sequence based on the lightcurves. Some of these stars that we discarded are not necessarily misclassified but could be the result of some pollution as we use larger aperture than the ones { optimized} for planetary research. As we want to measure the rotation and level of magnetic activity of the stars, we prefer to remove any star whose lightcurve has some pollution to be sure that our study is not biased.

In this work we are interested in studying solar-like stars for which acoustic modes have not been detected. Thus we removed from the sample the solar-like stars that have detection of acoustic modes from \citet{2011Sci...332..213C}, leading to a sample of 1,380 stars.  In addition, we re-analyzed the latest data release (DR25) with the best calibration available. Indeed, the latest lightcurves processed by the \kep\ Science office corrected some smearing effect that non-negligibly improved the noise at high-frequency leading to a higher signal-to-noise ratio and allowing us to detect acoustic (p) modes in additional stars. { In order to search for acoustic modes in main-sequence stars, we use short-cadence data. Given that none of these 1,380 stars had oscillations detected, they were dropped off the short-cadence program to give priority to other targets and only have short-cadence data available for one month.} We performed an asteroseismic analysis with the A2Z pipeline \citep{2010A&A...511A..46M} and the FliPer  \citep{2018A&A...620A..38B} to look for signature of p modes. We confirmed the detection of oscillations in 42 stars and still have 92 candidates. The detailed analysis of these new detections will be presented in Mathur et al. (in prep.).

Finally, we also ran the FliPer$_\textnormal{Class}$ tool for classification \citep{2019A&A...624A..79B} and did additional visual checks of the lightcurves. We found a sample of stars that presented peaks that did not look like the normal rotation behaviour we usually find so we decided to remove them from the sample.

After removing these stars, we end up with { 1,014} stars that are supposed to be main-sequence stars and that do not show any signature of oscillations. Figure 1 represents the Hertzsprung-Russel (HR) Diagram where the effective temperature and the surface gravity come from the DR 25 \kep\ stellar properties catalog \citep{2017ApJS..229...30M}. The grey symbols represent the KASC WG1 stars observed in short cadence during the survey phase. The stars without oscillations detected are represented by black circles. We superimposed the stars with p-mode detections (red circles) and the new detection candidates (blue crosses). We will focus on the stars without detection of oscillations (or the non-oscillating stars) in the following sections. { We also represent the HR Diagram of each sample in different panels for a better clarity.}

{
\subsection{Predicted mode amplitude}
There is a clear relation between the amplitude of the modes and the luminosity of the stars. \citet{1995A&A...293...87K} and \citet{1991ApJ...371..396B} derived scaling relations based on the Sun that related the maximum amplitude of the solar-like oscillations and the stellar parameters. These relations have been used in particular to predict the amplitude of the modes for a given solar-like star \citep{2011ApJ...732...54C, 2016ApJ...830..138C}. {  We used the relation of \citep{2016ApJ...830..138C} where we removed the multiplicative factor of 0.85 that was taking into account the redder response of TESS compared to \kep, leading to lower amplitudes.} We computed the predicted maximum amplitude of the modes for the 1,014  stars without detection of acoustic modes and for the 529 stars with detected oscillations using the {  following relation:

\begin{equation}
A_{\rm max} = {\rm A}_{\rm max, \odot}\,\times\,\beta\big{(}\frac{R}{R_\odot}\big{)}^2\big{(}\frac{T_{\rm eff}}{{\rm T}_{\rm eff, \odot}}\big{)}^{0.5} {\rm ppm},
\end{equation}}

where $T_\text{eff}$ is the effective temperature, $R$ is the stellar radius, and $\odot$ denotes the solar values (${\rm T}_{\rm eff,\odot}$\,=\,5777\,K; ${\rm R}_\odot=6.955\,\times\,10^{10}$cm). $A_{\rm max, \odot}$ is the root-mean-square maximum amplitude for the Sun (${\rm A}_{\max, \odot}$=2.5\,ppm). The factor $\beta$ is a correction that depends on the temperature of the star and is defined as:

\begin{equation}
\beta = 1 - \exp{\big{(}-\frac{T_{\rm red}-T_{\rm eff}}{1550}\big{)}}
\end{equation}

with $T_{\rm red}$ = 8907$(L/L_{\odot})^{-0.093}$\,K.

Here, we used the temperatures from the DR25 \kep\ star properties catalog \citep{2017ApJS..229...30M} and radii from \citet{2018ApJ...866...99B}) that incorporated DR1 $Gaia$ parallaxes to the previous catalog. We then computed the noise level for each star by taking the mean value of the power spectrum density in the frequency range 5,000\,$\mu$Hz to the Nyquist frequency for the short cadence ($\sim$\,8300\,$\mu$Hz). Figure~\ref{Amax_pred_Noise} represents the ratio of the predicted amplitude and the high-frequency noise as a function of the effective temperature ({  top} panel) and as a function of the surface gravity ({  bottom} panel). As expected, we can see that a large fraction of the stars with detected oscillations  (red symbols) have a ratio larger than 1. However {  9} stars, which are on the main sequence, are just below. We checked the metallicity of those stars and there is no systematic behaviour. We also compared the predicted amplitudes with the observed amplitudes for that sample of stars. We found that in average, the predicted amplitudes are {  slightly overestimated} compared to the observed ones. {  However}, two of the most studied stars (KIC 8006161 and KIC 10644253) are predicted to have a ratio close to 1 but their observed signal-to-noise ratio (SNR) is much higher. 

Given that around {  280} non-oscillating stars have a predicted maximum amplitude lower than the high-frequency noise in the power spectrum, the non detection of the modes for that sample can be explained by the too low signal-to-noise ratio. For the remaining {  734} stars with $A_{\rm max, pred}/{\rm Noise} \ge 1$, we should have detected the acoustic modes.  As mentioned above, some stars with $A_{\rm max, pred}/{\rm Noise}$ slightly below 1 have modes detected so for the next sections, we will look into stars that {  follow the criteria:

\begin{equation}
\frac{A_{\rm max, pred}}{\rm Noise} > 0.94.
\end{equation}}

In addition, for the comparison with the stars where we detected modes, we select stars that are in the same region of the HR Diagram with:

\begin{equation}
T_{\rm eff} \le 6,800\,K, 
\end{equation}

and

\begin{equation}
\log g \le 4.3\,{\rm dex}
\end{equation}

This yields a sample of 470 non-oscillating stars and 397 oscillating stars that are represented in Figure~\ref{HRD_select}.

}

\section{Data Analysis}\label{sec:data_an}

In this section, we present the methodology used to measure the surface rotation of the stars and how we determine the level of magnetic activity using the photometric data of \textit{Kepler}.

\subsection{Measuring rotation period}\label{sec:prot}
The surface rotation period can be measured from photometric observations through the signature of the regular passage of spots inducing periodic modulations of the luminosity. Several methods were thus developed providing the measurements of the rotation of thousands of main-sequence stars observed by {\it Kepler} \citep[e.g.][]{2013A&A...560A...4R,2014A&A...572A..34G,2014ApJS..211...24M}. These methods are based either on Lomb-Scargle periodograms \citep{2013A&A...557L..10N}, auto-correlation functions \citep[ACF,][]{2014ApJS..211...24M}, or wavelet-based analysis \citep{1998BAMS...79...61T,2010A&A...511A..46M}. 
In a blind hare-and-hounds exercise performed by most of the principal teams working on this topic with the {\it Kepler} data,  \citet{2015MNRAS.450.3211A} showed that the best
combination of completeness and reliability was obtained with the analysis technique developed by \citet{2014A&A...572A..34G} and \citet{2017A&A...605A.111C} using a combination of period-search methods such as ACF and wavelets. In this work, we thus applied the pipeline from \citet{2014A&A...572A..34G} to extract the surface rotation period of the { 1,014} stars. We analyzed the three datasets corresponding to the three filters as described in Section 2.1. First we selected the stars where all the methods and all the filters agreed within 2\,$\sigma$, where $\sigma = \sqrt{\sigma_1^2 + \sigma_2^2}$ with $\sigma_i$ being the uncertainty obtained by each method (except the ACF for which we do not compute any error). We also applied the criteria on the height of the ACF and the { Composite Spectrum} as described in \citet{2017A&A...605A.111C}. We then visually checked the remaining stars that were not selected according to all these criteria.  

We also analyzed the PDC-msMAP lightcurves to check for pollution as we use customized apertures with KADACS, { which are more likely to be polluted}. The crowding (that is the percentage of flux coming from the targeted star) of the stars in our sample is above 80\%, i.e. they { should have very little} polluted light. However, one thing to keep in mind when studying the surface rotation periods of the stars is that the PDC-msMAP lightcurves can be filtered with a { 21}-day filter, which means that we can only detect rotation shorter than $\sim$\,20 days \citep[see the discussion in][]{2013ASPC..479..129G}. This also implies that for a star rotating slower, the PDC-msMAP analysis can lead to the detection of a harmonic. When the rotation period found with PDC-msMAP and  KADACS agree within 2\,$\sigma$, we keep the rotation period found with the KADACS lightcurve. When the rotation obtained with PDC-msMAP is a harmonic of the period from KADACS, we also keep the rotation period from KADACS. Finally, if the rotation period (or a harmonic) is not found in the PDC-msMAP lightcurves, either the signal has been filtered out in these time series or the signal detected in the KADACS lightcurves could originate from a polluting star. Therefore, we discard this star.

\subsection{Photometric activity level: Sph}
\citet{2010Sci...329.1032G} showed in the case of the F-star HD\,49933 observed by CoRoT that the variability of the lightcurve, estimated from the standard deviation of the observations, is associated to the existence of spots rotating on the surface of the star, and thus can be used as proxy of stellar activity. Since then, several metrics were developed and applied 
to study the stellar activity of the {\it Kepler} targets \citep{2010ApJ...713L.155B,2011ApJ...732L...5C,2014JSWSC...4A..15M, 2014ApJ...783..123C}. However, the variability in the lightcurves can have different origins with different time scales such as magnetic activity but also convective motions, oscillations, or stellar companions. By taking into account the rotation of the star in its computation as it is done in \citet{2014JSWSC...4A..15M}, the derived activity metric, so-called $S_\mathrm{ph}$, is thus an actual proxy for magnetic variability. { We first compute} the standard deviation of  subseries of length  5$\,\times P_{\rm rot}$ \citep{2014JSWSC...4A..15M}. { Then we calculate the average of these standard deviations to derive the final $<S_\mathrm{ph}>$.} 

{ By comparing the $<S_{\rm ph}>$ with other classical proxies for the solar magnetic activity, such as sunspots or Ca II K-line emission, \citet{2017A&A...608A..87S} found a high correlation between the different magnetic proxies, showing the validity of our photometric proxy. }
 Nevertheless, in the stellar case, the estimated proxy, $S_\mathrm{ph}$, will depend on the inclination angle of the star in respect to the line of sight, { on the moment of the magnetic cycle when the star is observed, and the position of the active latitudes. These are also parameters that affect the spectroscopic observations for other classical proxies of stellar magnetic activity. So the $S_{\rm ph}$ is as good as a proxy as the Mount Wilson S-index. We just note that given the limitations listed above, it} should be considered as a lower limit of the stellar photospheric activity.


\section{Results of the analysis}\label{sec:results}

We performed the analysis of the rotation as described in the previous section { for the full sample of 1,014 stars}. The comparison of the different filters provided a list of reliable rotation periods for 412 stars. We then checked the remaining stars visually adding { 278} stars to the sample. We decided to use the following priority for the values of the rotation period. First, we select the value from the wavelet analysis (when the values agree within 2$\sigma$), then we select the value from the composite spectrum, and finally the one computed with the ACF. The $<S_{\rm ph}>$ is the one computed with the { finally selected} rotation period. The analysis of the PDC lightcurves and the comparison of their results with the KADACS results flagged 38 stars that we checked visually. We found that 6 could be due to pollution and discarded them. We end up with a list of 684 stars with rotation periods measured. Table 1 provides the list of stars with their { measured rotation periods, $S_{\rm ph}$ values, and their fundamental stellar parameters (effective temperature, surface gravity, and metallicity). Table 2 gives the list of stars without detection of rotation periods.}



We classified the stars in a similar way to \citet{2014A&A...572A..34G}: hot stars  with $T_{\rm eff} \ge 6250$\,K, dwarfs with $T_{\rm eff} < 6250$\,K and $\log g > 4$\,dex, and subgiants with $T_{\rm eff} \le 6250$\,K and $\log g \le 4.0$\,dex.   There are 313 stars from the studied sample coded as { hot stars, 327 stars as dwarfs, and 44 stars as subgiants. They are showed in a HR Diagram in the left panel of Figure~\ref{HRD_color} where we color-coded the different spectral types defined above. The right panel shows the same diagram for the stars selected following criteria (3), (4), and (5)}. 

Figure~\ref{Prot_hist} represents the distribution of the rotation periods measured for the different categories of stars mentioned above, as well as for the full sample of stars with reliable measurements (black line).  The left panel shows the hot stars and we clearly see that they are mostly rapid rotators as seen in \citet{2014A&A...572A..34G}. This is in agreement with the theory as hot stars (i.e massive stars) have thinner outer convection zones leading to a smaller braking due to stellar winds. The { mass limit between the hot stars and the cool stars is set to 1.3\,$M_\odot$} that is also called the Kraft break \citep{1967ApJ...150..551K}. So around the subgiant phase, stars with higher masses than 1.3$M_\odot$ will not undergo braking while cooler stars with lower masses will slow down. However in the middle panel of Figure~\ref{Prot_hist}, we observe many dwarfs with small rotation periods. One explanation is that the fundamental stellar parameters (temperature and gravity) are less reliable and some stars in the dwarf sample might actually be hotter stars. { Another plausible explanation is that these fast rotating cool stars are younger and have not yet slowed down.} Comparing the rotation period of the subgiants (right panel of Figure~\ref{Prot_hist}) to the ones from the hot stars and the dwarfs, we find that the number of subgiants is much smaller. Indeed, stellar evolution theory predicts that when a star evolves on the subgiant branch, it slows down and its magnetic activity decreases, i.e. less spots are present at their surfaces. Since our measurements depend on the passage of spots on the stellar surfaces, we do not expect to measure the surface rotation in many of these more evolved stars. Our results corroborate this theory. We also notice that the subgiants with a measured rotation period are rather fast rotators ({ shorter than} 30 days).

In Figure~\ref{Sph_hist}, we can see a histogram representation of the photospheric magnetic proxy, $<S_{\rm ph}>$, where the black dashed lines represent the range of the magnetic index of the Sun between minimum and maximum activity ({ $<S_{\rm ph, \odot, min}>$=67.4\,ppm and $<S_{\rm ph, \odot, max}>$=314.5\,ppm} respectively).  The hot stars (left panel of Figure~\ref{Sph_hist}) have in general a { similar level of magnetic activity to the Sun, while the dwarfs appear to be more active (middle panel of Figure~\ref{Sph_hist}), which is not what we expected}. For these two categories, we see that the distribution peaks very close to the maximum activity of the Sun. Given the uncertainties on the magnetic index, we could say that stars where no acoustic modes have been detected have slightly larger magnetic activity levels than the Sun.  As the sample of subgiants is very small we cannot conclude on their magnetic activity. Here again, this can be explained by the fact that subgiants are less active than main-sequence stars. 

{ We then represent the distribution of the rotation periods (Figure~\ref{Prot_histo_select}) and the magnetic activity proxy  (Figure~\ref{Sph_histo_select}) for the 323 non-oscillating stars as selected in Section 2.3 (criteria (3), (4), and (5)).  We compare those distributions with the ones for the oscillating stars in the same region of the HR Diagram. For the hot stars (left panels), the distribution of the oscillating and non-oscillating stars are very similar, peaking at a rotation period below 5 days and at an $<S_{\rm ph}>$ value of 300\,ppm. However, we can note that there are more non-oscillating stars with $<S_{\rm ph}> >$ larger than 300\,ppm. For the cool stars (middle panels), the rotation distributions are similar though there seems to be a bimodality. Given the small number of stars, it is not clear whether it is just a selection effect.  Concerning the magnetic activity of the cool dwarfs, non-oscillating stars are more active than the oscillating ones. Finally for the subgiants, given the small number of non-oscillating subgiants, it is harder to compare the distributions. The rotation periods of the non-oscillating stars are in general shorter and somewhat more active compared to the oscillating ones, which could suggest that these are young subgiants.}

\section{Discussion}\label{sec:discussion}

In Figure~\ref{Sph_Prot}, we show the magnetic index, $S_{\rm ph}$, as a function of the rotation period, $P_{\rm rot}$.  We compare the non-oscillating stars studied in this work (black symbols) and the { 310} oscillating stars with measured rotation periods from \citet{2014A&A...572A..34G} (red symbols).  The  dashed lines delimit the region of minimum and maximum levels of magnetic activity of the Sun based on the VIRGO \citep[Variability of solar IRradiance and Gravity Oscillations][]{1995SoPh..162..101F} photometric observations \citep{2014A&A...572A..34G}.  { Figure~\ref{Sph_Prot_select} is the same as Figure~\ref{Sph_Prot} but for the stars {  with  $A_{\rm max, pred}/{\rm Noise} > 0.94$} and selected in the same region of the HR Diagram.} 

\subsection{Magnetic activity effect}

{ We first start the analysis of the full sample.} We can first notice that { 252 stars ($\sim$ 80\%)} with detected oscillations have an $<S_{\rm ph}>$ value that falls in the same range as the Sun between minimum and maximum activity.  Indeed, we expected that { cool dwarfs} with similar level of activity to the Sun have detection of p modes. However, we have 41 stars with a level of magnetic activity above the maximum of the Sun for which we have detected oscillations.  We were not expecting them to have detected modes as they have high magnetic activity levels. We checked the maximum amplitude of the modes obtained from A2Z but those stars do not have a systematic low amplitude. They could be either subgiants or metal-rich stars \citep{2010A&A...509A..15S,2013A&A...549A..12M} enhancing the mode amplitudes and compensating the magnetic effect. 

According to the DR25, 22 of these stars are subgiants so this could be a valid explanation for detecting their p modes (their intrinsic oscillations amplitudes are higher than during the main sequence). For the remaining 19 stars, the metallicity ([Fe/H]) obtained by the \kep\ Follow-up Observations Program \citep{2018ApJ...861..149F} also used in the DR25 stellar properties is super-solar for 12 stars. We still have 7 stars that have a high $S_{\rm ph}$ value, low [Fe/H] but for which we could detect solar-like oscillations. { This sounds reasonable given that for the Sun the amplitudes decrease by only around 12.5\% \citep[e.g.][]{2018SoPh..293..151K} so the amplitudes could still be large enough to be detected even for stars with a $<S_{\rm ph}>$ value above $<S_{\rm ph, \odot, max}>$.}

We see that below the $S_{\rm ph}$ at minimum solar activity, there are around 100 stars with rotation and magnetic activity detection whether they have detected p modes or not. We can consider that the threshold of rotation and magnetic activity detectability is $\sim$\,20-30 ppm. We can also derive a threshold on $<S_{\rm ph}>$ of 2,000\,ppm above which the probability of non detection of p modes is of 98.3\%.

Regarding the stars without oscillations detected, as expected they have larger magnetic indexes compared to the oscillating sample. We find that 47.7\% of the non-oscillating stars have an $<S_{\rm ph}>$ larger than $<S_{\rm ph, \odot, max}>$. These stars agree and confirm that a high level of magnetic activity suppress the oscillations and can prevent us from detecting solar-like oscillations as also shown by \citet{2011ApJ...732L...5C}.

{ If we now look at the 323 stars selected as described in Section 2.3 with a measured rotation period, we find that 103 stars have $<S_{\rm ph}> > <S_{\rm ph, \odot, max}>$. So 68\% of the stars where we expect high enough amplitudes  to detect the acoustic modes have a level of activity similar to the Sun or lower.  This is quite interesting and unexpected given the claim of \citet{2011ApJ...732L...5C} and the anti-correlation found between the Mount Wilson S-index and maximum amplitude of the modes for a small sample of stars by \citet{2014A&A...571A..35B}}.

{ In Figure~\ref{Sph_Prot_cat} and \ref{Sph_Prot_cat_select}, we have divided the sample in the three categories for the full sample of stars and for the selected sample from criteria (3), (4), and (5) respectively in order to see where the different stars are situated in the HR Diagram. We can see that the non-oscillating hot stars (top left panels) are mostly concentrated in the same region as the oscillating stars. The cut from Section 2.3 removes some of the very active stars. The sample of cool dwarfs (top right panels) without detection of p modes  contains many more active stars compared to the cool dwarfs with detected oscillations. The cut of the sample removes many of the high $<S_{\rm ph}>$ stars as many were very cool dwarfs with low expected amplitudes of the modes but some stars with high activity still remain. Finally the cut of the subgiants (bottom panels) removes some very active subgiants as well.}


{ Since we find a large number of stars without detection of p modes and with magnetic indexes in the same range as the oscillating stars, we propose to investigate three explanations for these stars}: 

\begin{itemize}
    \item  the inclination angle affects the measurement of $<S_{\rm ph}>$ as with a low inclination angle { not all the active regions are} observed. As a consequence, the value we measure is just a lower limit of the real magnetic activity proxy of the star;
    \item correlation between the acoustic-mode amplitudes and the metallicity. Indeed it has been shown that metal poor stars have lower mode amplitudes \citep{2010A&A...509A..15S,2010A&A...518A..53M}. { However metallicity also has an effect on the magnetic activity. \citet{2018ApJ...852...46K} showed how the magnetic activity level ({ obtained through the} S-index and $<S_{\rm ph}>$) is larger for a super-metallic solar-like star. So the abundance can have two different effects that could also counter balance each other};
    \item the stars could be in a binary system \citep{Schonhut-Stasik2019}. Indeed a close-in companion could suppress p modes through tidal effects as it has been observed in red giants belonging to binary systems \citep{2014ApJ...785....5G}.
\end{itemize}

\begin{figure}[htbp]
\begin{center}
\includegraphics[width=20cm]{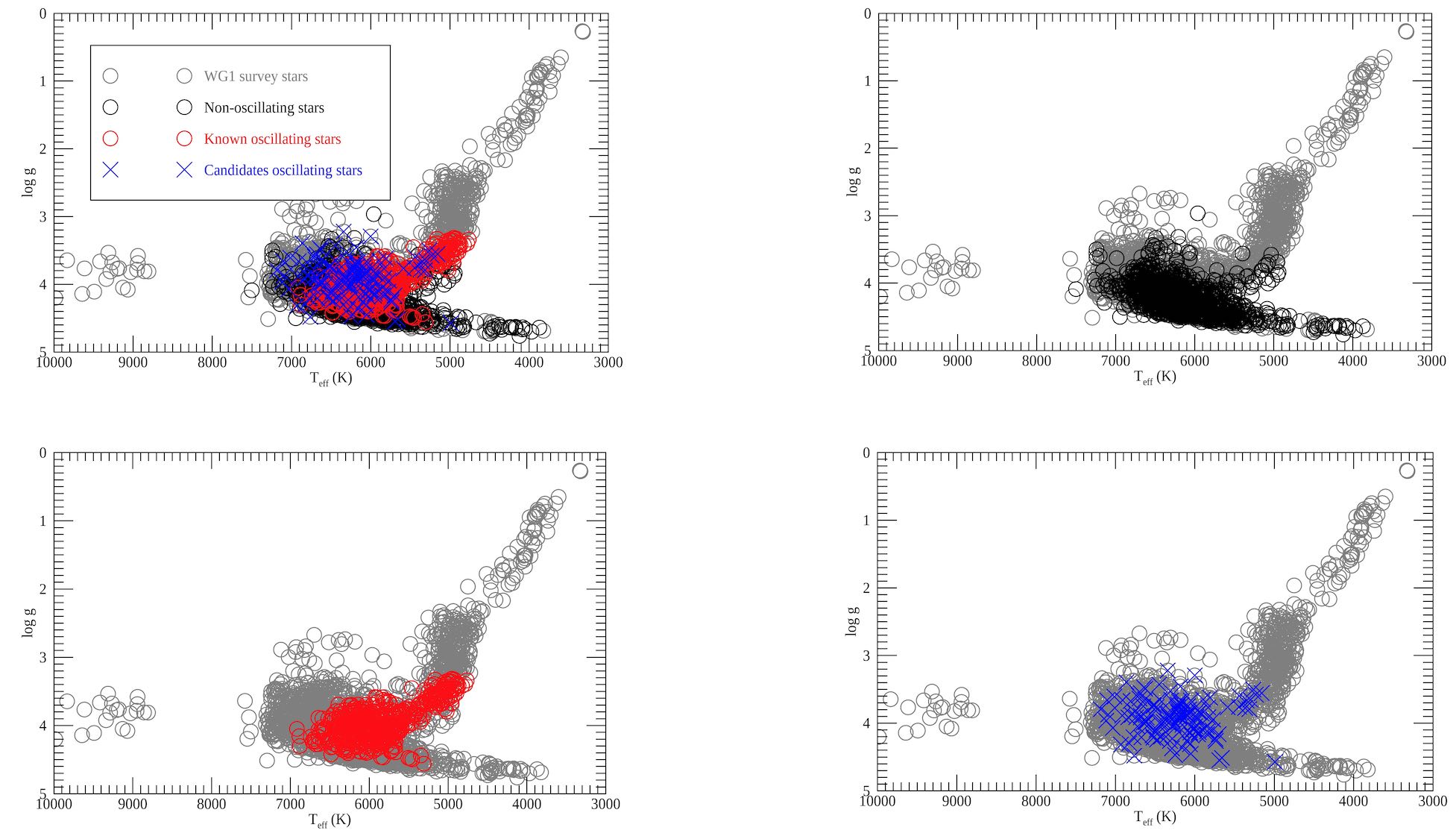}
\caption{Hertzsprung-Russel Diagram for the different samples of stars mentioned in the paper. The grey circles represent KASC WG1 stars that were observed in short cadence during the survey phase. The red circles  are the main-sequence stars with known oscillations from literature. The black circles correspond to the final set of { 1,014} stars obtained as described in Section 2.2. { The blue crosses are the new candidates with detection of oscillations.}}
\label{HRD}
\end{center}
\end{figure}
\begin{figure}[htbp]
\begin{center}
\includegraphics[width=12cm]{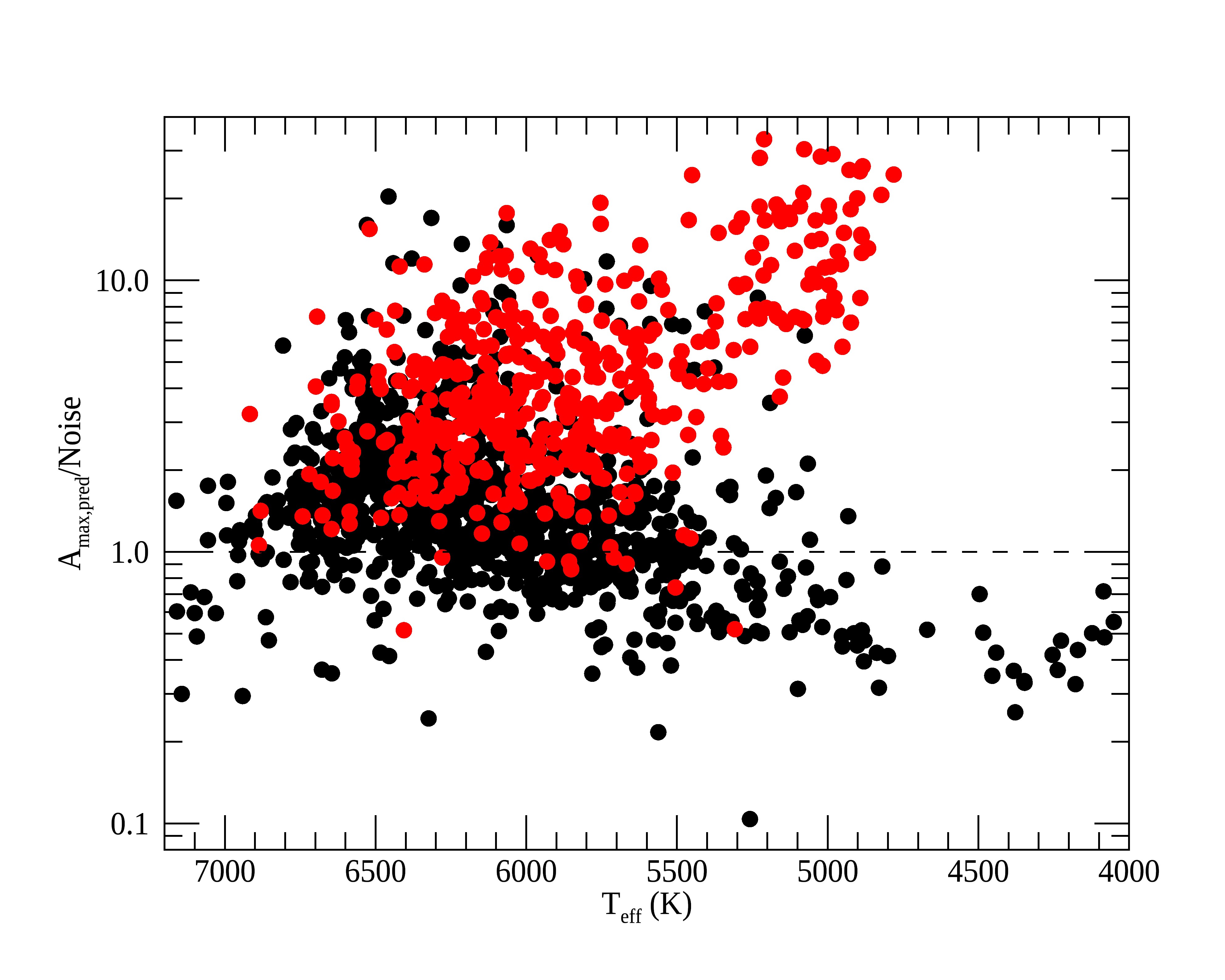}
\includegraphics[width=12cm]{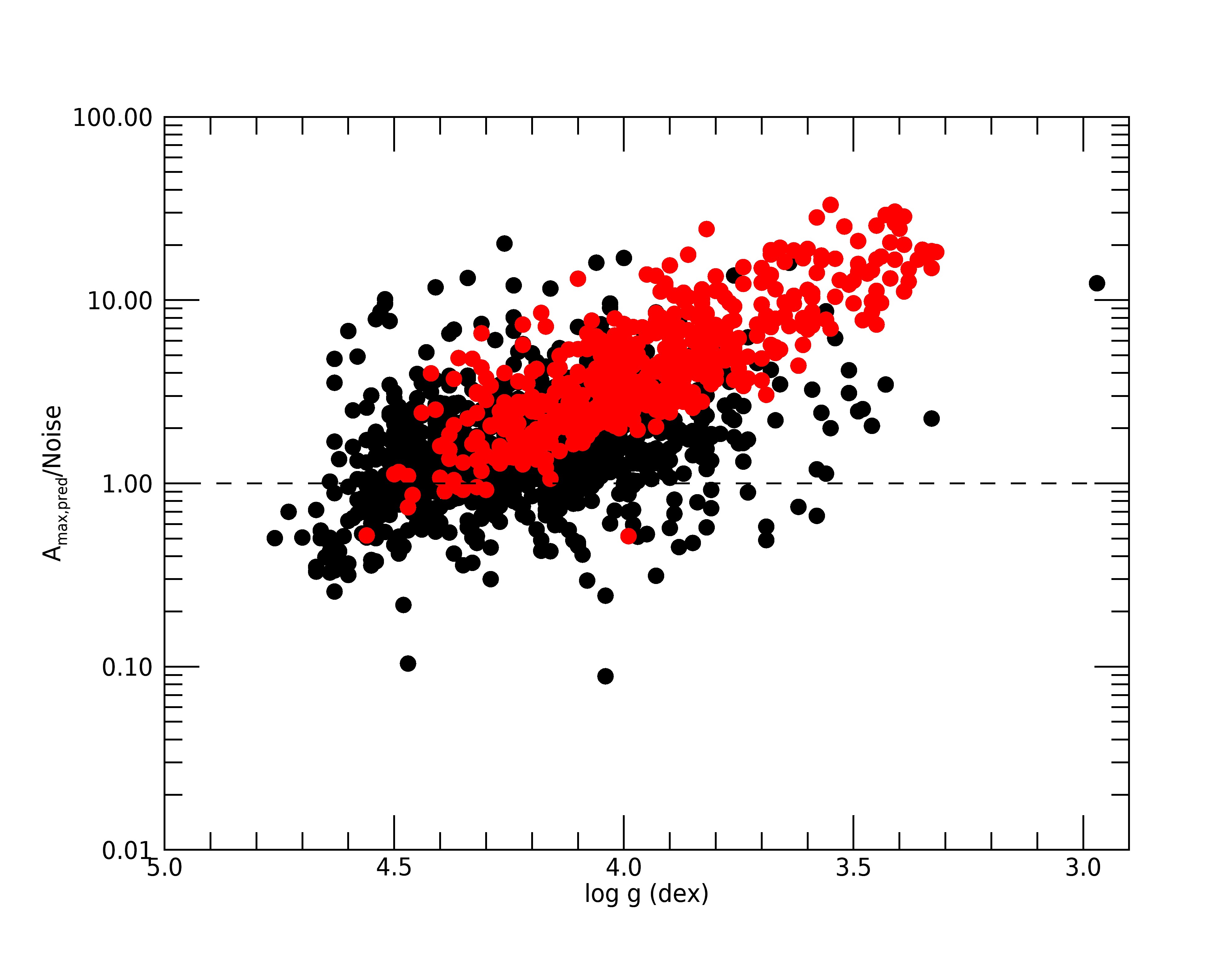}
\caption{{ Ratio of the predicted maximum amplitude and the noise at high frequency as a function of the effective temperature {  (top panel) and as a function of surface gravity (bottom panel)} for the stars without detected oscillations (black circles) and the stars with detected oscillations (red circles).}}
\label{Amax_pred_Noise}
\end{center}
\end{figure}
\begin{figure}[htbp]
\begin{center}
\includegraphics[width=12cm]{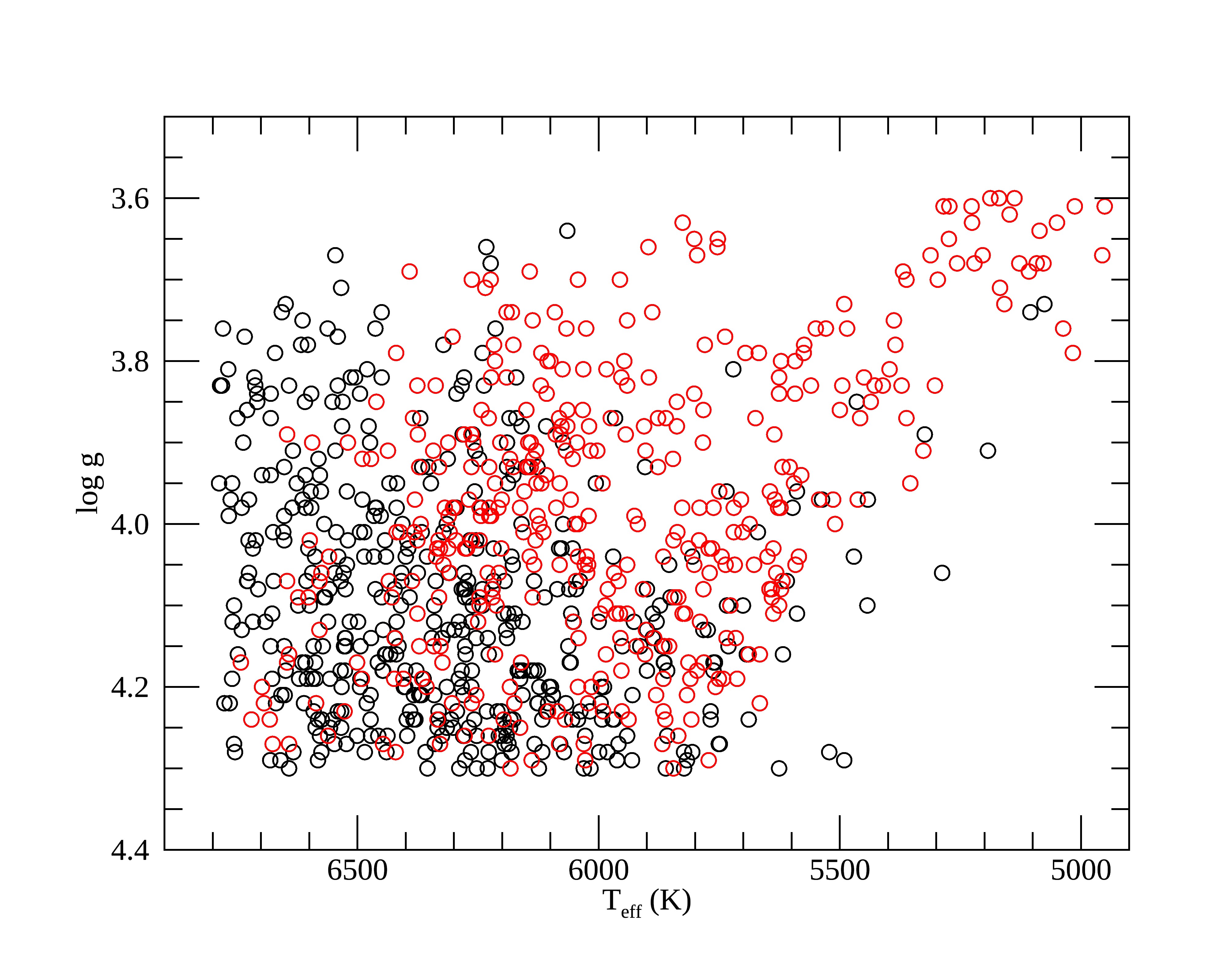}
\caption{{ Hertzprung-Russel Diagram for the stars selected as described in Section 2.3, with $A_{\rm max,pred}/{\rm Noise} > 0.94$, $T_{\rm eff} < $6,800\,K, and $\log g <$4.3\,dex. Stars without detected oscillations are represented with black circles and stars with detected oscillations in red circles.}}
\label{HRD_select}
\end{center}
\end{figure}
\begin{figure}[htbp]
\begin{center}
\includegraphics[width=10cm]{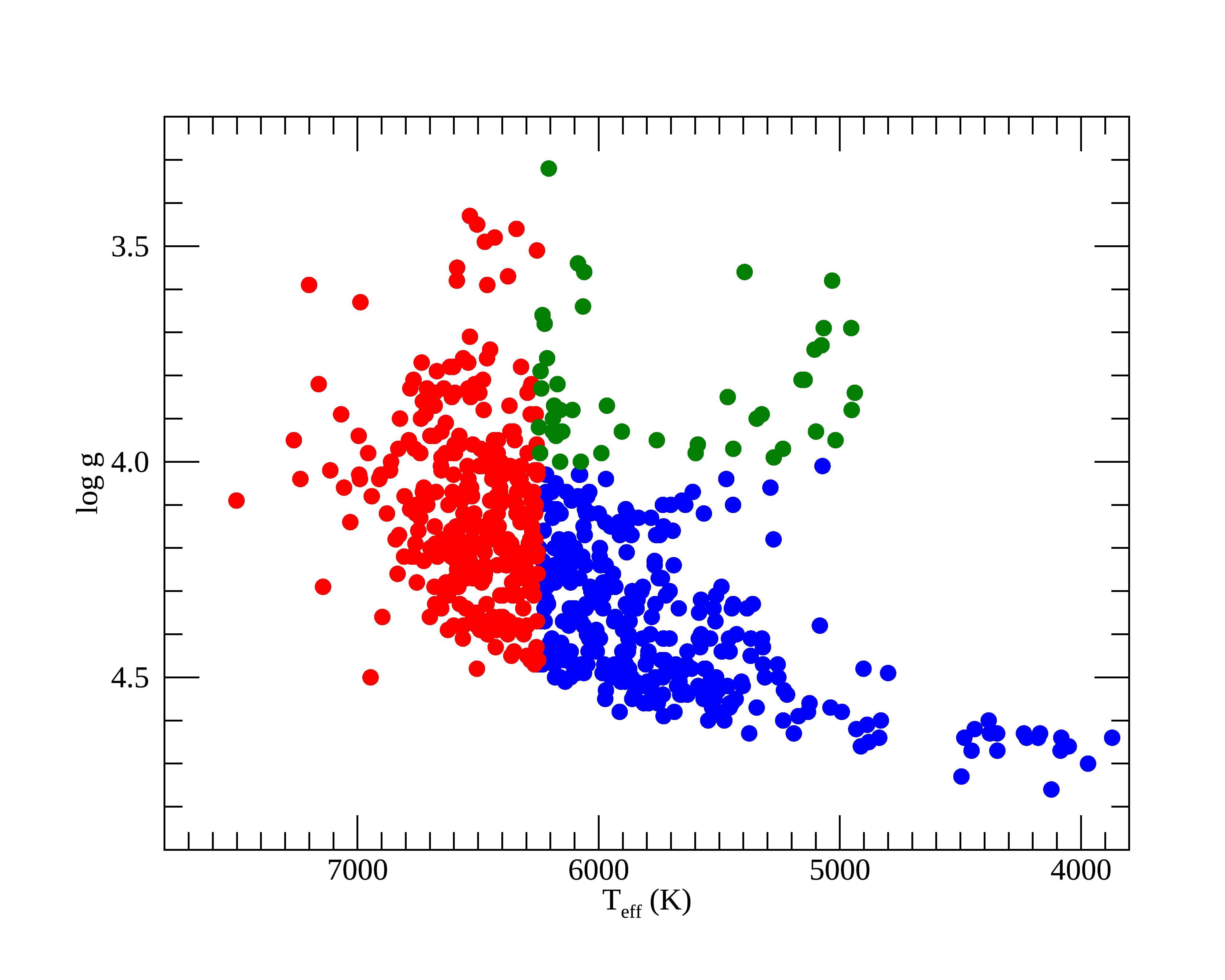}
\includegraphics[width=10cm]{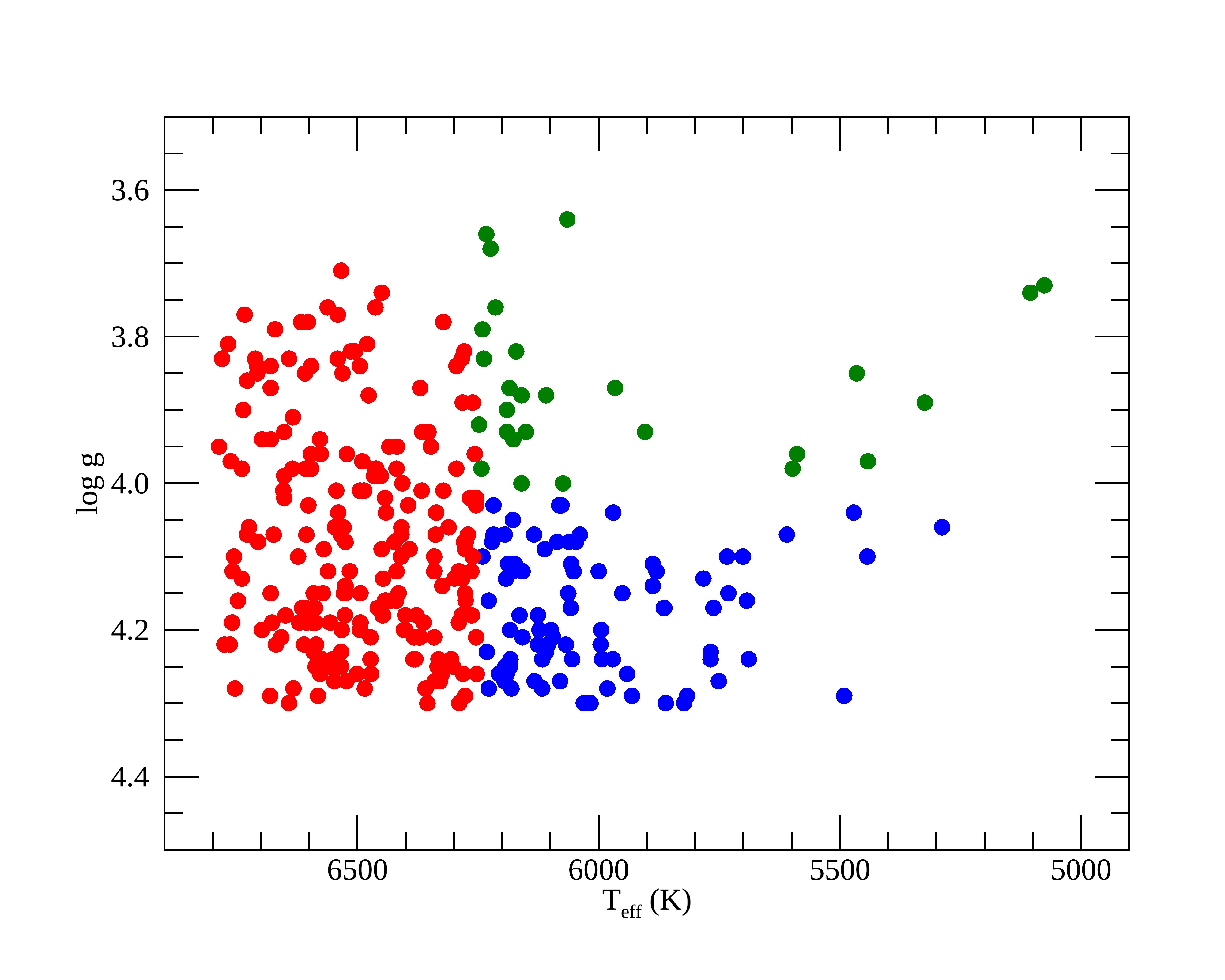}
\caption{{ HR Diagram showing the hot dwarfs (red symbols), cool dwarfs (blue symbols), and subgiants (green symbols) as described in Section 4. Top panel: all the non oscillating stars with a measurement of rotation periods. Bottom panel: stars selected from the cut described in section 2.3 with a measurement of rotation period.}}
\label{HRD_color}
\end{center}
\end{figure}
\begin{figure}[htbp]
\begin{center}
\includegraphics[width=12cm]{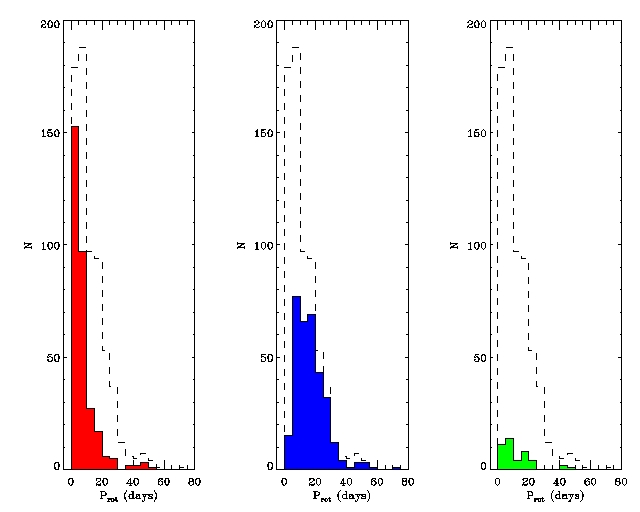}
\caption{Distribution of the rotation periods for { the full sample and} the three categories of stars defined in Section 4: hot stars (left panel), cool dwarfs (middle panel), and subgiants (right panel). The black dash line in each panel represents the distribution for the full sample of stars with reliable rotation periods.}
\label{Prot_hist}
\end{center}
\end{figure}
\begin{figure}[htbp]
\begin{center}
\includegraphics[width=12cm]{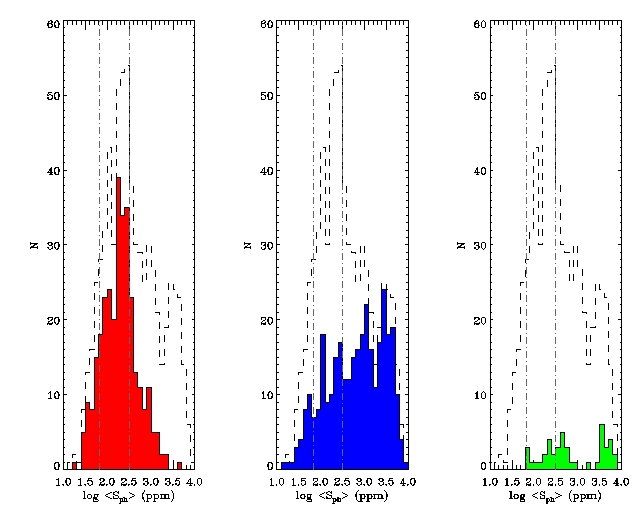}
\caption{Distribution of the $<S_{\rm ph}>$ for { the full sample and} the three categories of stars as described in Section 4. Same legend as Figure~\ref{Prot_hist}. The grey dashed-dotted lines represent the values at minimum and maximum magnetic activity for the Sun.}
\label{Sph_hist}
\end{center}
\end{figure}
\begin{figure}[htbp]
\begin{center}
\includegraphics[width=12cm]{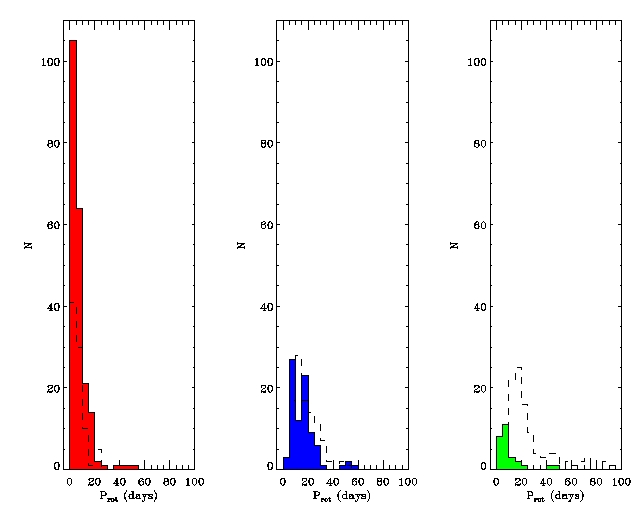}
\caption{{ Rotation distribution for the non-oscillating stars selected as in bottom panel of Figure~\ref{HRD_color} with the same color code. The dash lines in each panel correspond to the rotation distribution of the oscillating stars in each spectral type.}}
\label{Prot_histo_select}
\end{center}
\end{figure}
\begin{figure}[htbp]
\begin{center}
\includegraphics[width=12cm]{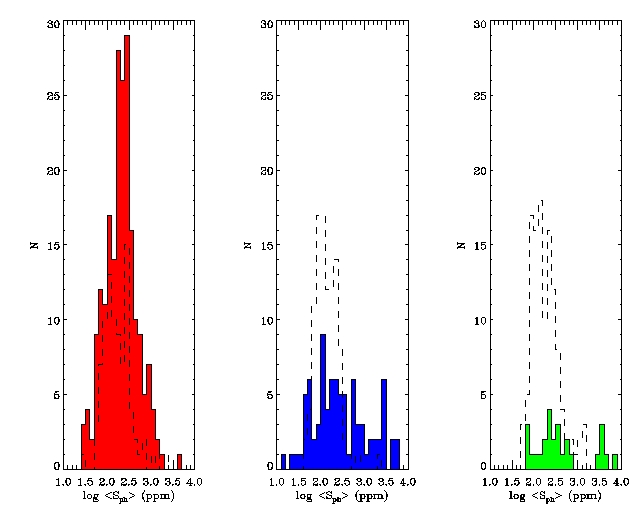}
\caption{{ Distribution of the $<S_{\rm ph}>$ for the non-oscillating stars selected as in bottom panel of Figure~\ref{HRD_color} with the same color code. The dash lines in each panel correspond to the magnetic activity distribution of the oscillating stars in each spectral type.}}
\label{Sph_histo_select}
\end{center}
\end{figure}
\begin{figure}[htbp]
\begin{center}
\includegraphics[width=12cm]{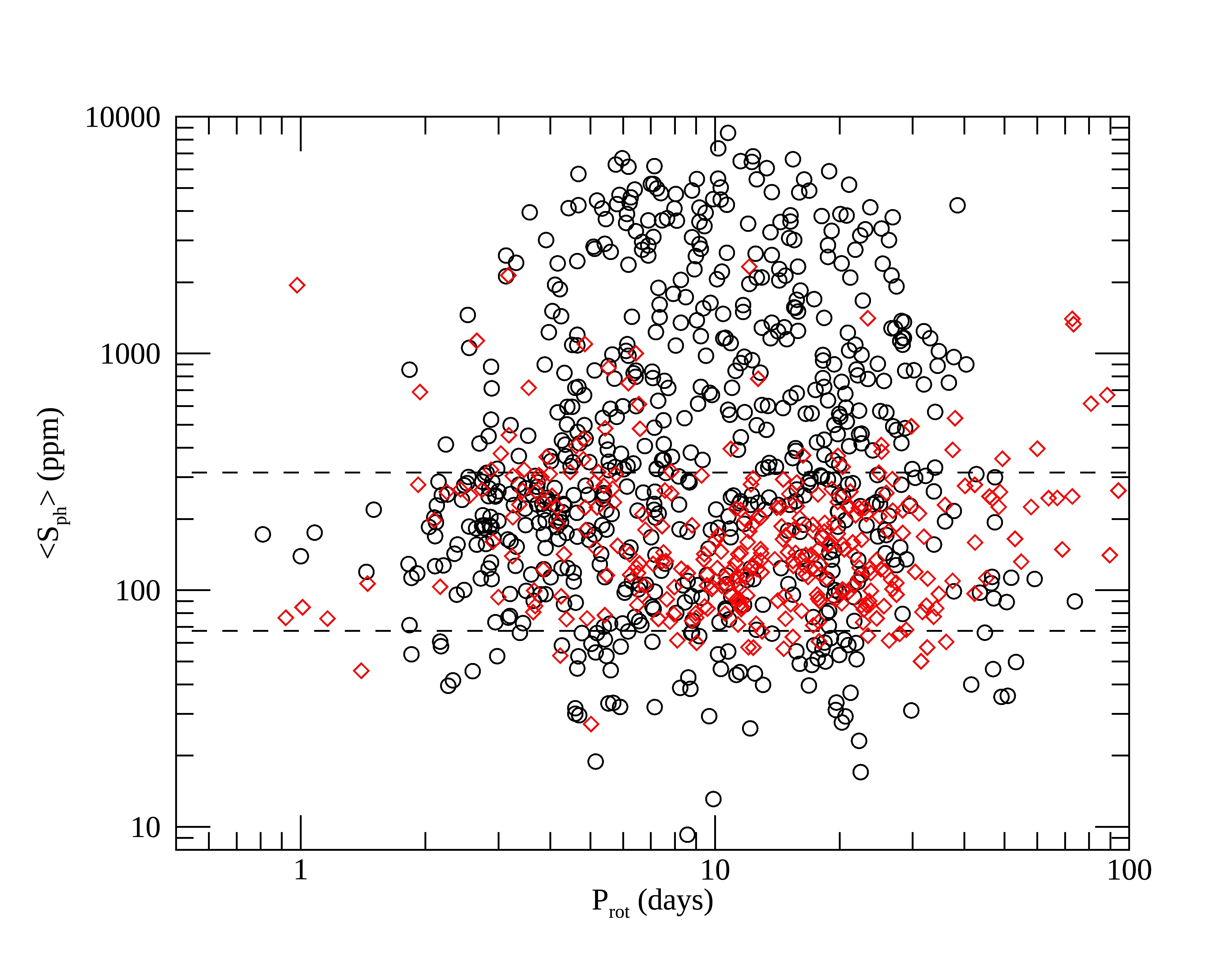}
\caption{Proxy of the magnetic activity, $<S_{\rm ph}>$ as a function of the rotation period $P_{\rm rot}$ {  for} the full sample. The black symbols represent the non-oscillating stars while the red symbols represent the stars with detected oscillations from Garc\'{i}a et al. (2014). The dash lines are the $<S_{\rm ph}>$ values at minimum and maximum activity of the Sun.}
\label{Sph_Prot}
\end{center}
\end{figure}
\begin{figure}[htbp]
\begin{center}
\includegraphics[width=12cm]{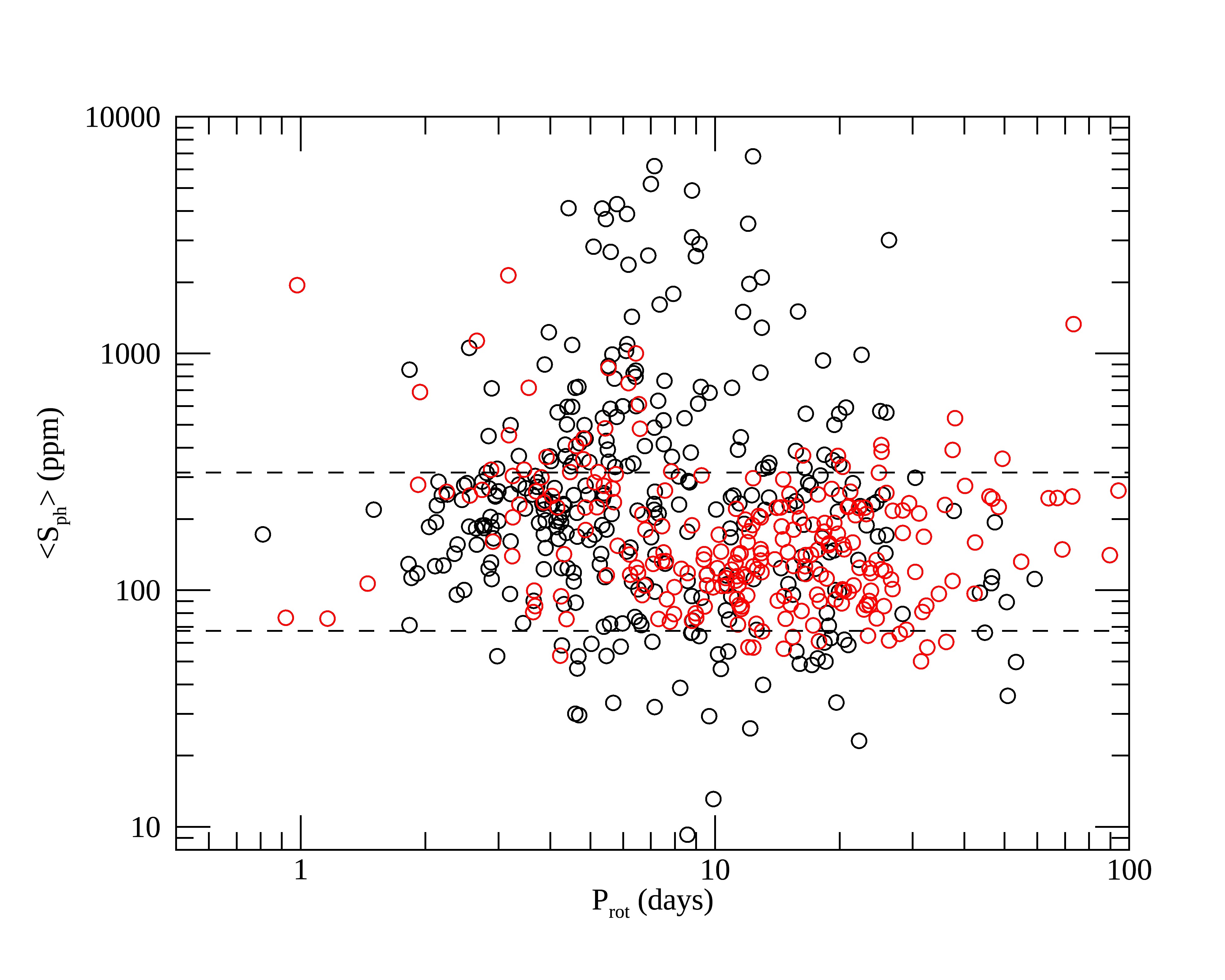}
\caption{{ Same as Figure~\ref{Sph_Prot} but for stars selected as in Section 2.3.}}
\label{Sph_Prot_select}
\end{center}
\end{figure}
\begin{figure}[htbp]
\begin{center}
\includegraphics[width=18cm]{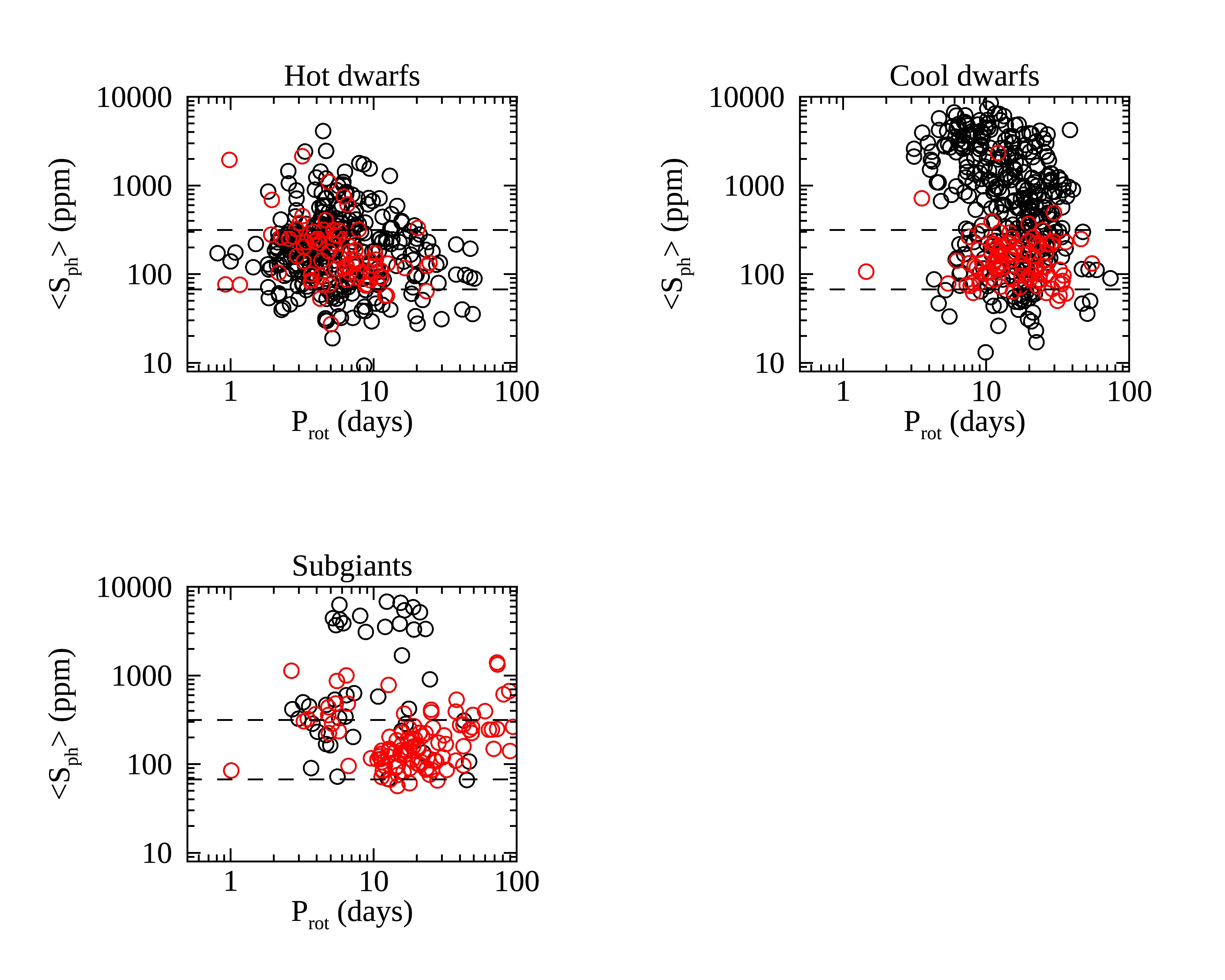}
\caption{Same as Figure~\ref{Sph_Prot} but separated into the three categories of stars: hot dwarfs, cool dwarfs, and subgiants.}
\label{Sph_Prot_cat}
\end{center}
\end{figure}
\begin{figure}[htbp]
\begin{center}
\includegraphics[width=18cm]{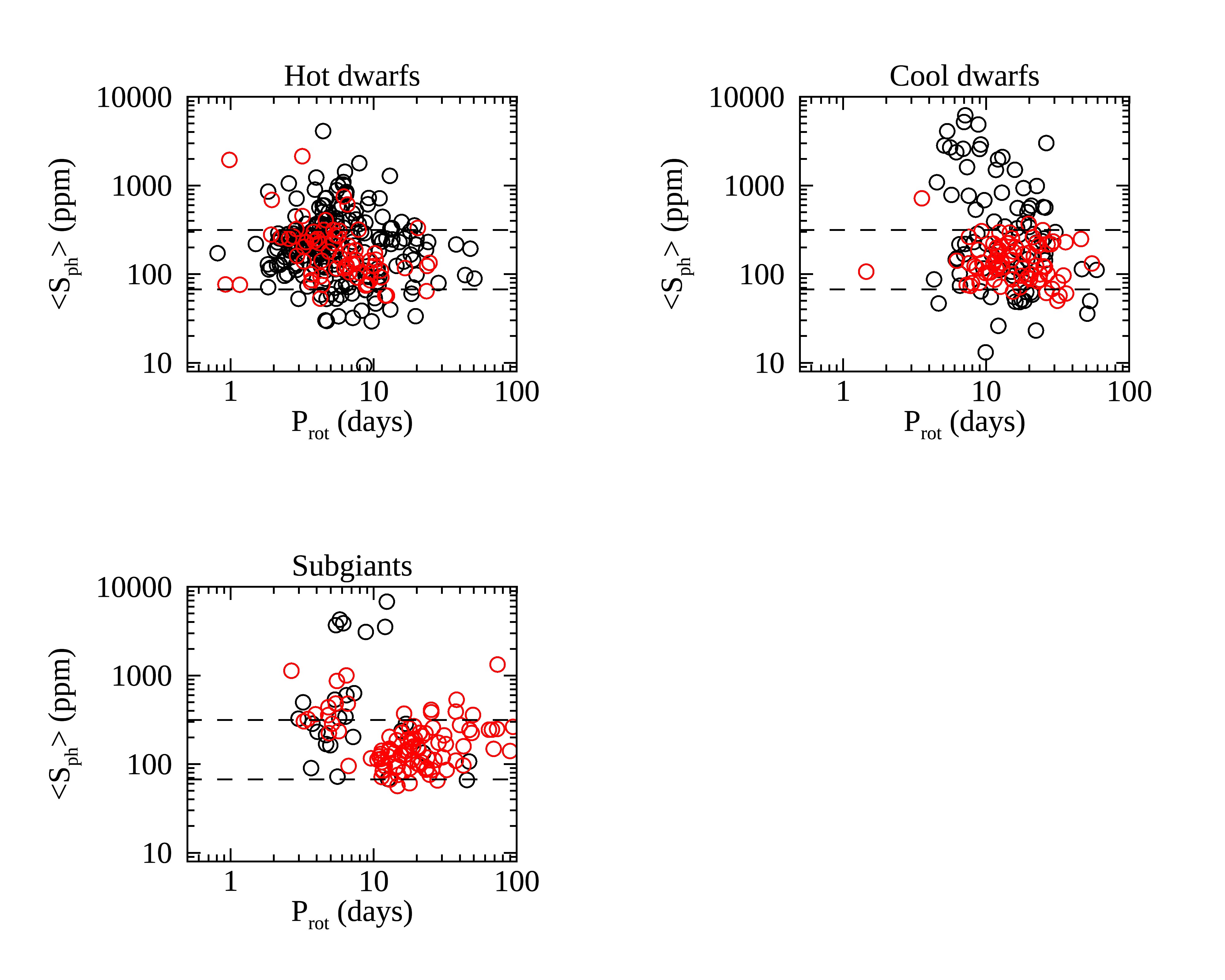}
\caption{Same as Figure~\ref{Sph_Prot_select} but separated into the three categories of stars: hot dwarfs, cool dwarfs, and subgiants.}
\label{Sph_Prot_cat_select}
\end{center}
\end{figure}
\begin{figure}[htbp]
\begin{center}
\includegraphics[width=12cm]{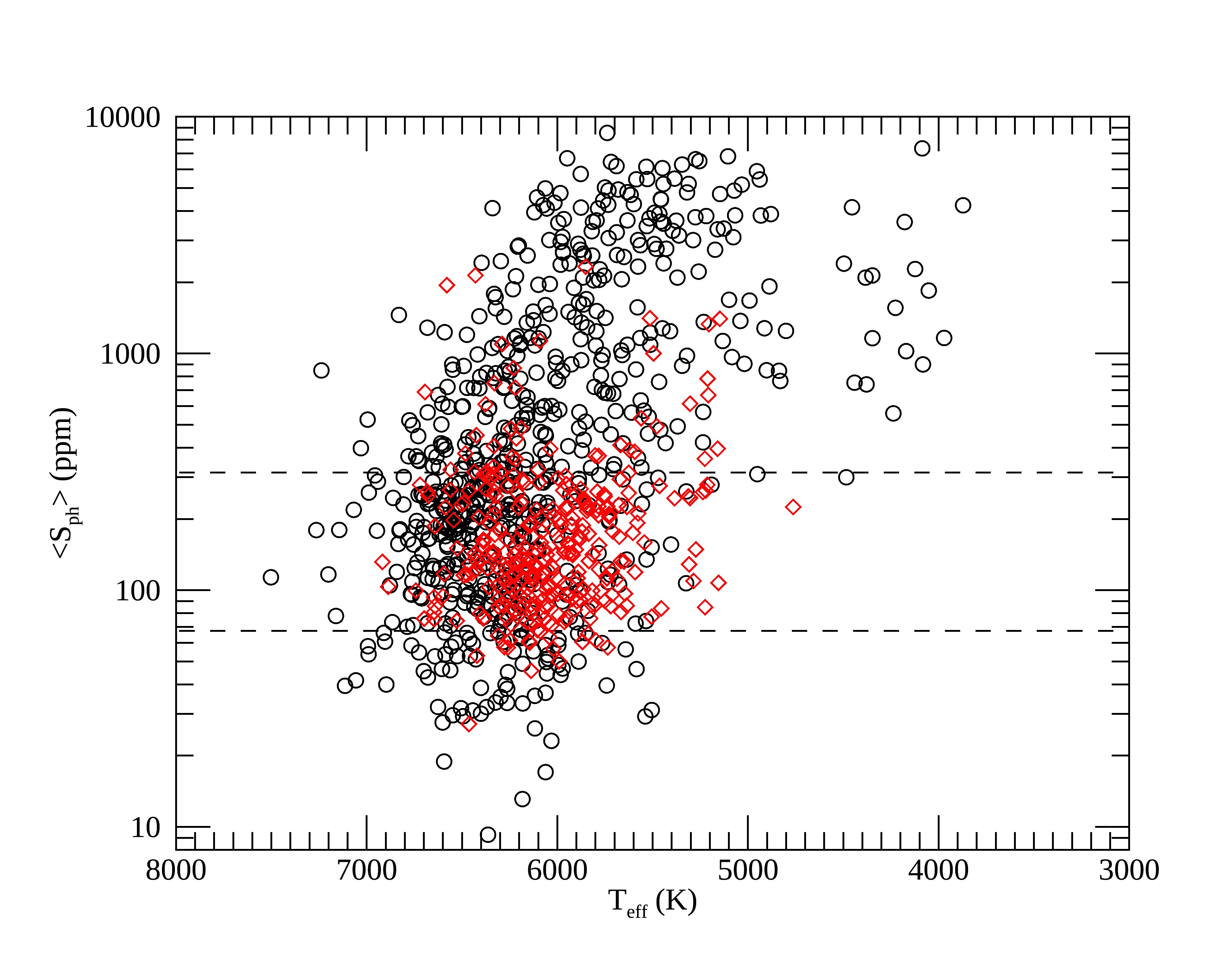}
\caption{$<S_{\rm ph}>$  as a function of the effective temperature from the DR25 stellar properties catalog. Same legend as Figure~\ref{Sph_Prot}.}
\label{Sph_Teff}
\end{center}
\end{figure}

{ In Figure~\ref{Sph_Teff}}, we also represented the $S_{\rm ph}$ as a function of effective temperature. We do not see any correlation between the magnetic activity level and the $T_{\rm eff}$. From Figure~\ref{HRD}, the non-oscillating stars had a temperature range between 3,800\,K and 7,500\,K. { We note that stars with very low levels of magnetic activity are rather hot stars with temperatures above 6,000\,K.}


\subsection{Metalicity effect}

\begin{figure}[htbp]
\begin{center}
\includegraphics[width=12cm]{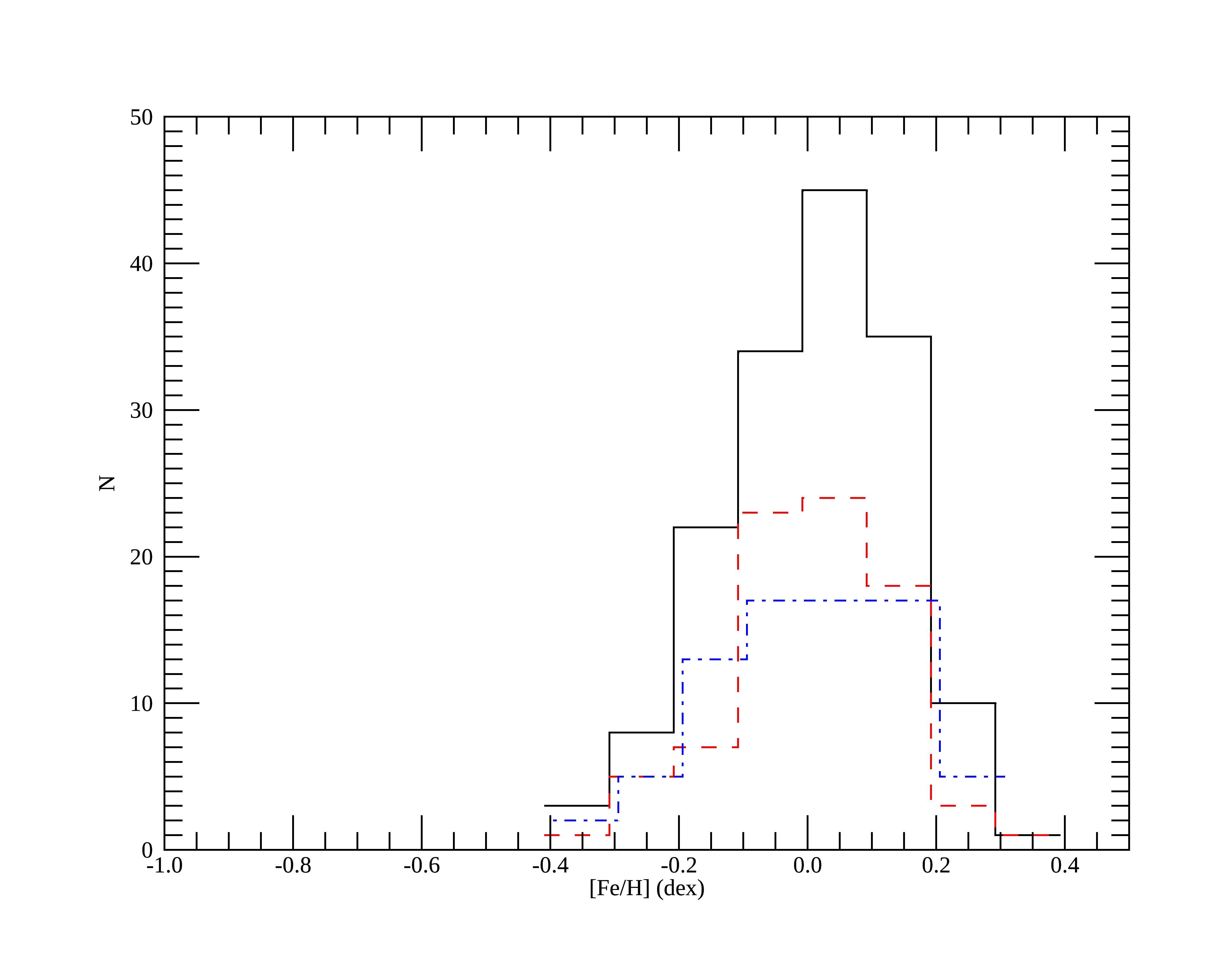}
\includegraphics[width=12cm]{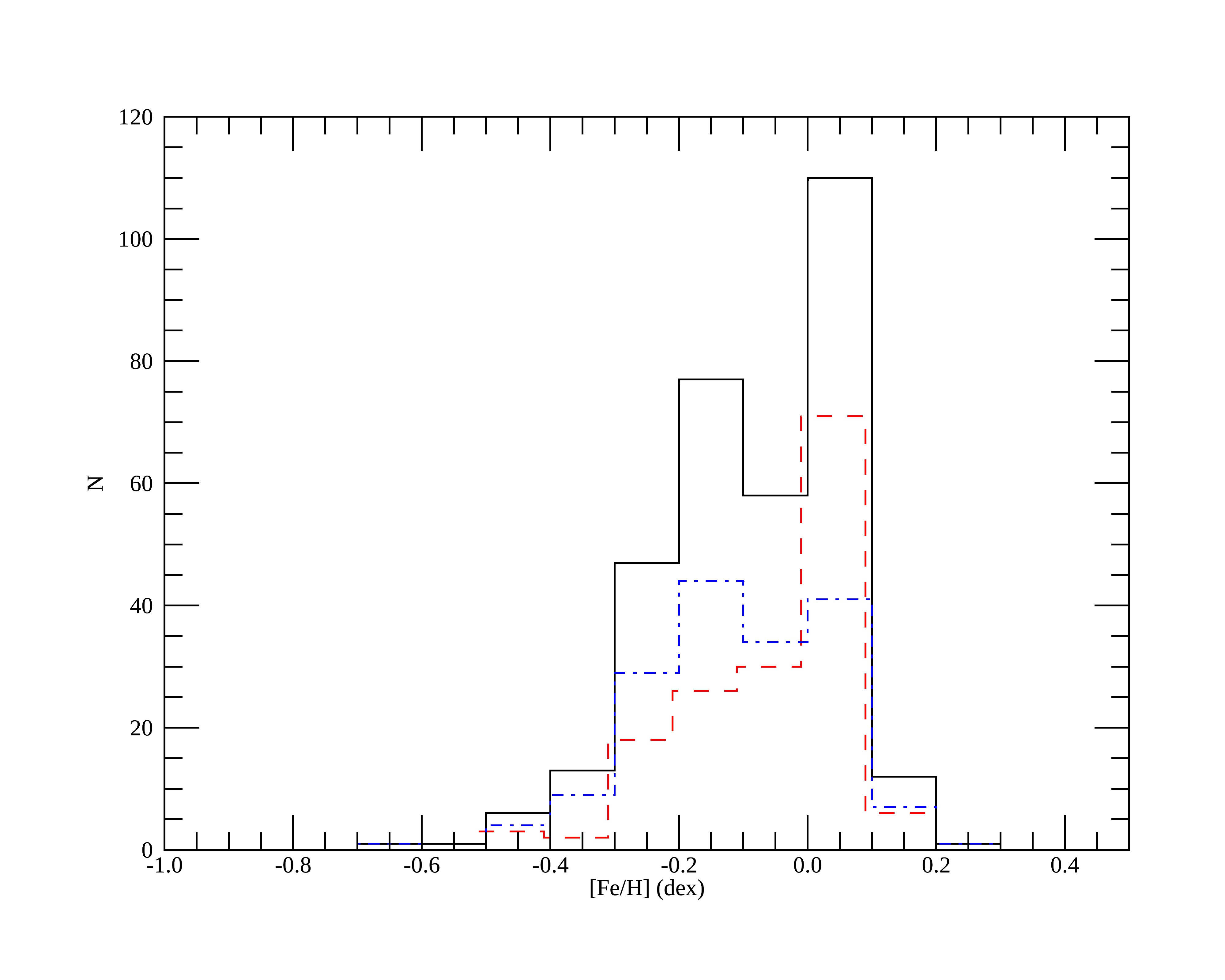}
\caption{Histogram of metallicity for the non-oscillating stars using APOGEE DR14 for { 158} stars (top) and LAMOST DR2 for { 326 stars} (bottom panel). The full sample in each panel is represented with the black solid line. Blue dot-dash line corresponds to the least active stars { ($S_{\rm ph} <S_{\rm ph, \odot, max}$)} while red dash line corresponds to the most active stars ({ $S_{\rm ph} >S_{\rm ph, \odot, max}$}). }
\label{histo_FeH}
\end{center}
\end{figure}

Many of the non-oscillating stars have spectroscopic data obtained with the DR14 Apache Point Observatory Galactic Evolution Experiment survey \citep[APOGEE][]{2017AJ....154...94M,2018AJ....156..125H} and the DR2 Large Sky Area Multi-Object Fiber Spectroscopic Telescope survey \citep[LAMOST][]{2015ApJS..220...19D,2016yCat.5149....0L}, providing metallicity values more precise than photometric observations. After cross-checking our sample {  of stars with $A_{\rm max, pred} > 0.94$} with the sample of these surveys we find that { 158} stars have high-resolution spectroscopic observations from APOGEE and { 326} stars have low-resolution spectroscopic data from LAMOST. We note that there could be an overlap between these two samples. 
Figure~\ref{histo_FeH} shows the [Fe/H] histograms for APOGEE (top panel) and for LAMOST (lower panel) for all the non-oscillating stars that have a rotation and metallicity measurement (black solid line). { We can see some differences between the two surveys, where it seems that the stars with APOGEE metallicity have an average of 0.05\,dex while the ones with LAMOST observations have an average metallicity around -0.03\,dex. We compared the metallicity from both surveys for a common sample of around 6,000 stars and found a trend where at low APOGEE [Fe/H], the LAMOST values are larger while at high APOGEE [Fe/H], the LAMOST values are smaller. We then did the comparison for the common sample of non-oscillating stars and in general LAMOST has lower [Fe/H] values compared to APOGEE, which can explain the difference in the histograms from the two surveys. Given that the usual spectroscopic uncertainties for metallicity is around 0.15\,dex, we can still say that the two surveys are in agreement.}


We then focus on the least active stars (with $S_{\rm ph}$ smaller than the maximum activity value for the Sun) that are represented with the blue dot-dash line while the stars more active are represented with a red line. In the LAMOST sample, we clearly see that the average metallicity of the stars is close to the solar one while the non-oscillating not too active stars have in general a sub-solar metallicity. { We count 106 sub-solar metallic stars and 64 super-solar metallic stars.} This agrees with the theory of \citet{2010A&A...509A..15S} a lower metallicity will lead to a less opaque convection zone and hence less energy for the excitation of the modes. As a consequence metal-poor stars are expected to have smaller acoustic-mode amplitudes.

However for the stars with APOGEE metallicity, it seems that the very active stars have both super and sub-solar metallicity.  { We find that 24 stars have sub-solar metallicity while 52 have super-solar metallicity. } So the low metallicity only confirms the non detection of the modes in a smaller sample of stars. { As mentioned earlier, high metallicity could also lead to a higher level of magnetic activity \citep{2018ApJ...852...46K}, which could explain the non-detection of the modes for some of the super-metallic stars. This competing effect could be the source of the unclear influence of the metallicity and its study is out of the scope of this work.}

{ From this analysis, we note that the two surveys do not agree on the influence of metallicity. However as mentioned above within the typical uncertainties on [Fe/H] from spectroscopic observations, there is still some agreement. Besides the sample of stars being quite small the conclusions need to be taken cautiously. More high-resolution spectroscopic observations would be needed to better study that effect.}

\subsection{Inclination angle effect}

\begin{figure}[htbp]
\begin{center}
\includegraphics[width=12cm]{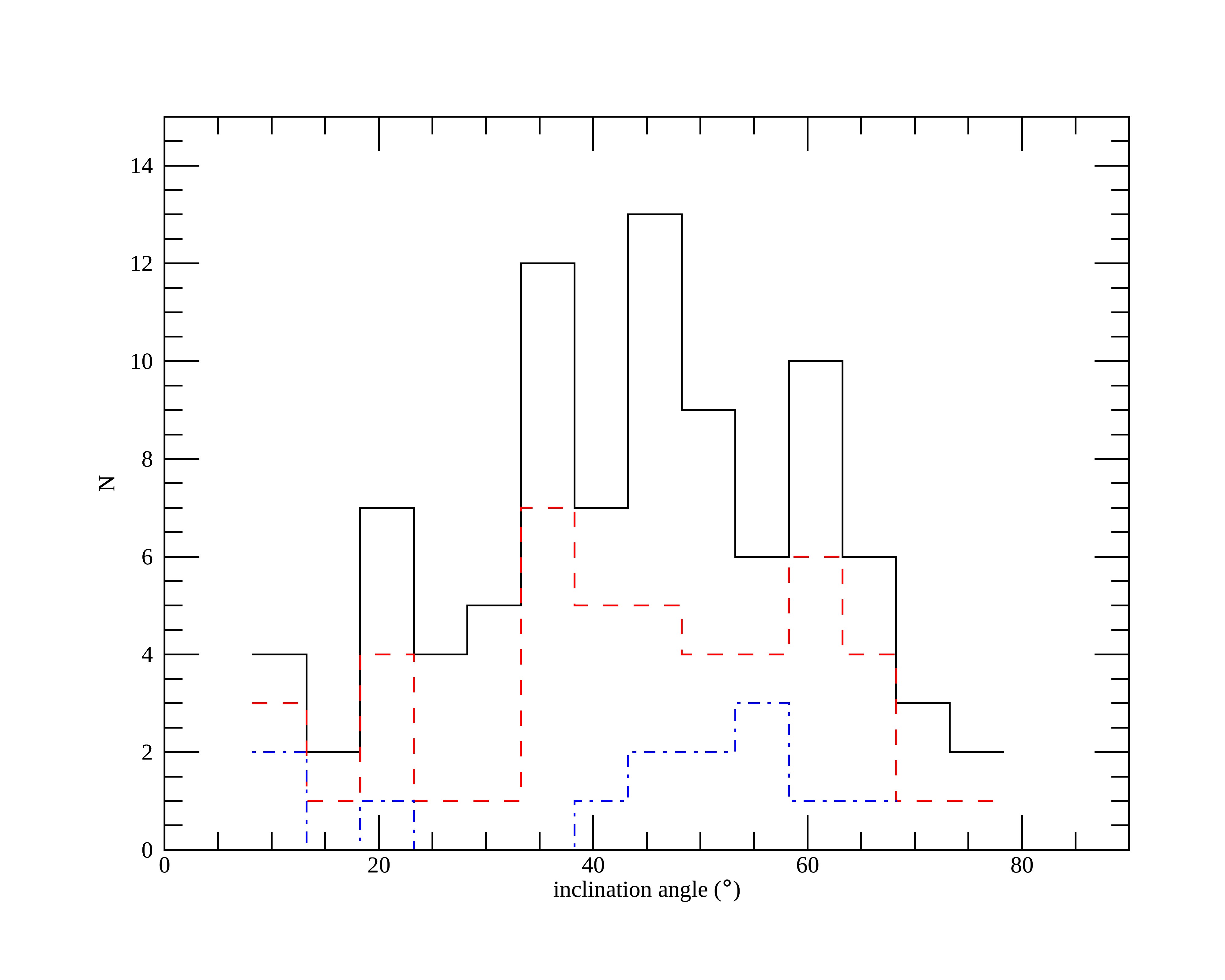}
\caption{Distribution of inclination angles for stars with $v \sin i$ from APOGEE { for stars with $A_{\rm max, pred}/{\rm Noise} > 0.94$}. The red dash line represents the least active stars ({ $<S_{\rm ph}> < S_{\rm ph, \odot, max}$} ) while the blue dot-dash line represents the least active stars with a super-solar metallicity.}
\label{inclination}
\end{center}
\end{figure}

The APOGEE spectroscopic observations also provide a measurement of the $v \sin i$ for some of our non-oscillating stars. By combining these values with the surface rotation periods measured with the \kep\ data and with estimates of the radii from the DR25 stellar properties, we can estimate the inclination angles of a subsample of { 162} stars { selected to follow criteria (3)}. We obtain a reliable inclination angle, $i$, for { 90} stars. For the other stars, the $\sin i$ is larger than 1. This could be due to a non-agreement between the rotation period and the $v \sin i$ or a less reliable radius, as the radii of these stars were mostly derived from isochrone fittings, hence less precise than asteroseismology. { The typical uncertainties of the inclination angles are around 25 degrees.} Figure~\ref{inclination} represents the distribution of the inclination angles for those stars. We can see that they peak around an angle of 45\,$\deg$. We then study the sample of stars with an $<S_{\rm ph}>$ lower than { $<S_{\rm ph, \odot, max}>$}  as we want to understand why the modes of these stars are not detected. The distribution (dashed red line in Figure~\ref{inclination}) is slightly flatter but the median value is still around 45\,$\deg$. Since some of these stars also have a low metallicity that already explains the non-detection of the modes, we selected the stars with a metallicity larger than the Sun. That sample consists of 13 stars, among which 4 stars have a low inclination angle and 9 have an inclination angle above 45\,$\deg$ (blue dot-dash line). With this small number of stars, we cannot conclude that the low-inclination is responsible for the low $<S_{\rm ph}>$.

\subsection{Binarity effect}

We cross-checked our sample of stars with the stars flagged as binary with the {\it Gaia} catalog \citep{2018ApJ...866...99B}. Among our sample of non-oscillating stars with rotation periods, 614 stars are included in the aforementioned catalog. We found that 8 stars of our sample are potential binaries. KIC~4057892, KIC~4930560, KIC~8737920, and KIC~12470631 have a flag 1 standing for a binary candidate based on the Gaia radius. KIC~4276716, KIC~5564082, KIC~5617510, and KIC~8396660   have a flag 2, meaning that they are  AO-detected binaries  \citep{2018AJ....155..161Z}.

Recently, \citet{Schonhut-Stasik2019} studied how a stellar companion can suppress acoustic modes. They obtained ROBO-AO images and/or RV measurements for a sample of 327 stars (including red giants and solar-like stars without detection of modes and with low-amplitude modes). No clear difference between oscillating stars and non-oscillating stars was found, whether in terms of close or wide binarity. They conclude that binarity is not the only mechanism to inhibit oscillations. 

We cross-checked our sample of stars with reliable rotation periods and photometric magnetic index with their binary stars and found only three stars in common. Two of them have an $<S_{\rm ph}>$ below { $<S_{\rm ph, \odot, max}>$}. We would need a larger sample of stars with measured RV.

\subsection{HERMES observations}
Because non-oscillating stars are not typical targets of large-scale spectroscopic surveys, we 
started to pay close attention to those targets within our own spectroscopic observing programs. 
\cite{2016A&A...589A..27B,2017A&A...602A..63B} and \cite{2016A&A...596A..31S,2016A&A...589A.118S} performed spectroscopy of 18 oscillating solar analogues with the Hermes high-resolution spectrograph \citep{2011A&A...526A..69R, 2011PhDT.......203R}, mounted on the 1.2\,m Mercator telescope on La Palma, Canary Islands. We extended this program to cover a larger sample of oscillating as well as non-oscillating targets. Observing time was granted and will be executed in the coming semesters. 

As a proof of concept we can discuss the first three non-oscillating stars for which we obtained spectroscopic observations, KIC\,11498538, KIC\,7841024, KIC\,7898839. Note that the first of these stars (KIC~11498538) is not in our sample anymore as it was flagged by the FliPer$_{\rm Class}$ tool as a possible classical pulsator. According to the literature, that star is actually listed as an eruptive variable star, agreeing with the peculiar flag of FliPer$_{\rm Class}$. However being included in the preliminary work and still a candidate of a fast rotator without detection of oscillations, we will discuss the spectroscopic analysis of that star. { For the other two stars we have computed that the predicted amplitude of the modes is much larger than the noise.}

All stars have a significantly higher chromospheric activity than the Sun. KIC\,7898839  was chosen because of its solar-like rotation period of 28\,days. For the two other stars, a rotation period of $\sim$3\,days was derived from \kep\ photometry. Despite the similar rotation period KIC\,11498538 shows signs of significantly broader rotational broadening than KIC\,7841024, suggesting that the latter one is seen towards a pole-on inclination. We refrain from comparing the stars on the basis of the Mount-Wilson S-index \citep{1991ApJS...76..383D}, as such large rotational velocities are likely to broaden the emission peak beyond the narrow wavelength range, defined by the triangular filters.



Currently, these are only first diagnostics from a small sample of stars. We plan to have a final sample of several dozens of stars, all obtained with the same spectrograph, which will then be consistently analyzed.

\section{Conclusions}\label{sec:conclusions}

We have revisited the analysis of the 2,576 stars observed during the survey phase { of main-sequence solar-like pulsating stars} of the \kep\ mission with one month of short-cadence data and followed for four years in long cadence. After removing polluting stars (red giants, classical pulsators, { stars with already detected solar-like oscillations}, new detection of solar-like oscillations, probable pollution), we have a sample of { 1,014} main-sequence and subgiant solar-like stars. We measured the surface rotation period, $P_{\rm rot}$, and the photometric magnetic activity index, $<S_{\rm ph}>$ and obtained reliable values for 684 stars, representing a yield of 67.4\%. As expected the hot stars rotate in general faster than the cool dwarfs. Our sample of subgiants is too small to conclude anything about them. 

\begin{itemize}
   \item { We have computed the predicted amplitude of the modes for the 1,014 stars of our sample and compared it to the high-frequency noise in the power spectrum. The majority of  the stars with detected oscillations have a ratio between the predicted amplitude and the noise above 1, with only 16 stars below. We found that for 394 non-oscillating stars, that ratio is below 1.}
    \item We can provide a lower limit of detection of magnetic activity of 20-30\,ppm below which we do not detect any rotation periods.
    \item { If we discard the stars for which the predicted amplitude of the modes is smaller than the noise, we find that only 32\% (compared to 47.7\% for the full sample) of the stars have a magnetic activity larger than the Sun at maximum activity. While magnetic activity explains the non detection of the modes for some stars, this is not the case for a large fraction of the stars without detection of oscillations.}
    \item Based on our samples and the analysis presented here, we conclude that if a star has an $<S_{\rm ph}>$ larger than 2000\,ppm, it has 98.7\% probability that we will not detect oscillations. 
    \item We notice a sample of stars with a high magnetic activity level and for which p modes have been detected. We found that 12 stars have a high [Fe/H] and 22 stars are subgiants, which could explain the enhancement of the modes in spite of their large $<S_{\rm ph}>$ values.
    \item The study of the remaining stars without detection of modes shows that low metallicity could be an explanation but given that high-resolution spectroscopic observations for those stars were available only for only { 158} stars, we cannot firmly conclude.
    \item We also investigated the inclination angle using radius from the DR25 \kep\ stellar properties catalog and $v \sin i$ from APOGEE but no clear correlation was found. However the $v \sin i$ was available for a small sample of stars and the radius from the isochrone fitting from DR25 are not the most precise values, which could bias our analysis.
    \item We checked whether the stars belonged to binary systems and only a handful of them are flagged as binaries either by {\it Gaia} or by AO observations.
    \item The preliminary analysis of the spectroscopic observations from the HERMES instrument for three stars shows that they have a high level of chromospheric activity. 
\end{itemize}

While magnetic activity seems to explain the non-detection of oscillations in almost half of the sample and some hints that metallicity can also play a role, we still have stars for which there is no clear and firm explanation. We would need additional spectroscopic observations (that are underway). 

Understanding the non-detection of oscillations is very important given the future missions like the NASA Transiting Exoplanet Survey Satellite \citep[TESS,][]{2015JATIS...1a4003R} already launched and the ESA PLAnetary Transits and Oscillations of stars \citep[PLATO,][]{2014ExA....38..249R} coming up as they rely on the detection of solar-like oscillations to better characterize the stars, in particular the ones hosting planets.



\bibliographystyle{frontiersinSCNS_ENG_HUMS}
\bibliography{/Users/Savita/Documents/BIBLIO_sav}

\section*{Funding}
This paper includes data collected by the \emph{Kepler} mission. Funding for the \emph{Kepler} mission is provided by the NASA Science Mission directorate. Some of the data presented in this paper were obtained from the Mikulski Archive for Space Telescopes (MAST). STScI is operated by the Association of Universities for Research in Astronomy, Inc., under NASA contract NAS5-26555. Partly Based on observations obtained with the HERMES spectrograph on the Mercator Telescope, which is supported by the Research Foundation - Flanders (FWO), Belgium, the Research Council of KU Leuven, Belgium, the Fonds National de la Recherche Scientifique (F.R.S.-FNRS), Belgium, the Royal Observatory of Belgium, the Observatoire de Gen\`eve, Switzerland and the Th\"uringer Landessternwarte Tautenburg, Germany. S.M. acknowledges support by the National Aeronautics and Space Administration under Grant NNX15AF13G, by the National Science Foundation grant AST-1411685 and the Ramon y Cajal fellowship number RYC-2015-17697. 
R.A.G. acknowledges the support from PLATO and GOLF CNES grants. ARGS acknowledges the support from National Aeronautics and Space Administration under Grant NNX17AF27G.
P.G.B. acknowledges the support of the MINECO under the fellowship program 'Juan de la Cierva incorporacion' (IJCI-2015-26034).



\section*{Tables}

\begin{table}[h!]
\caption{Non-oscillating stars with { reliable rotation periods measured in this work}. { Surface gravity and effective temperatures come from the DR25 Stellar properties catalog (Mathur et al. 2017). Metallicity comes from the {\it Kepler} star properties DR25 (flag=0), APOGEE (flag=1), and from LAMOST (flag=2).}}
\begin{center}
\begin{tabular}{ccccccc}
\hline
KIC & $P_{\rm rot}$ (days) & $S_{\rm ph}$ (ppm) & $\log g$ & $T_{\rm eff}$ (K) & [Fe/H] (dex)& Flag\\
\hline
\hline
1164109 &     4.39\,$\pm$\,   0.47 &           501.62\,$\pm$\,          32.60
 &  3.984 &  6608 &  -0.260 &  0\\
1434277 &    10.32\,$\pm$\,   0.75 &          5026.23\,$\pm$\,         201.58
 &  4.499 &  5750 &   0.110 &  2\\
1576249 &    46.65\,$\pm$\,   5.69 &           113.81\,$\pm$\,           2.94
 &  4.072 &  6195 &  -0.120 &  0\\
1718828 &    47.36\,$\pm$\,   2.89 &           193.56\,$\pm$\,           2.52
 &  4.205 &  6372 &  -0.350 &  2\\
1724041 &    47.08\,$\pm$\,   6.46 &            92.38\,$\pm$\,           2.80
 &  4.097 &  6711 &  -0.240 &  0\\
1724355 &    51.88\,$\pm$\,  10.09 &           112.89\,$\pm$\,           6.42
 &  4.239 &  6127 &  -0.470 &  2\\
1864124 &     7.18\,$\pm$\,   0.56 &           141.04\,$\pm$\,           8.90
 &  4.042 &  6337 &  -0.900 &  0\\
1868918 &    12.48\,$\pm$\,   1.03 &            44.49\,$\pm$\,          33.10
 &  4.344 &  6054 &   0.038 &  1\\
2013883 &    59.09\,$\pm$\,   0.00 &           111.50\,$\pm$\,           2.61
 &  4.058 &  6225 &  -0.026 &  1\\
2308753 &     2.77\,$\pm$\,   0.34 &           182.90\,$\pm$\,           9.77
 &  4.150 &  6525 &  -0.020 &  2\\
2443534 &    10.86\,$\pm$\,   0.44 &           551.09\,$\pm$\,          22.96
 &  4.487 &  6091 &  -0.140 &  0\\
2445004 &    13.52\,$\pm$\,   2.67 &           344.65\,$\pm$\,          13.21
 &  4.169 &  6058 &  -0.180 &  0\\
2448426 &    13.31\,$\pm$\,   1.41 &           475.73\,$\pm$\,          18.09
 &  4.223 &  6257 &  -0.420 &  0\\
2571934 &     5.47\,$\pm$\,   0.39 &            52.75\,$\pm$\,           5.40
 &  4.167 &  6458 &   0.020 &  0\\
2578513 &     3.60\,$\pm$\,   0.22 &           116.55\,$\pm$\,          10.27
 &  3.594 &  7201 &  -0.840 &  0\\
2580928 &    22.42\,$\pm$\,   1.50 &           226.53\,$\pm$\,           7.20
 &  4.116 &  6052 &  -0.300 &  0\\
2709654 &     4.68\,$\pm$\,   0.45 &            52.55\,$\pm$\,          10.16
 &  4.303 &  6642 &  -0.240 &  0\\
2718678 &    24.52\,$\pm$\,   1.97 &           235.31\,$\pm$\,           7.18
 &  4.205 &  6095 &  -0.140 &  0\\
2722192 &    13.61\,$\pm$\,   1.06 &          3250.04\,$\pm$\,         114.00
 &  4.482 &  5688 &   0.170 &  2\\
2837133 &    22.42\,$\pm$\,   1.80 &          3151.36\,$\pm$\,          86.29
 &  4.334 &  5361 &   0.308 &  1\\
\hline
\end{tabular}
\end{center}
\label{default}
\end{table}%

\newpage

\begin{table}[h]
\caption{{ KIC numbers of the non-oscillating stars} without measured rotation periods. }
\begin{center}
\begin{tabular}{c}
\hline
KIC \\
\hline
\hline
     1849587\\
     2158850\\
     2443543\\
     2445385\\
     2450979\\
     2696938\\
     2849969\\
     2987466\\
     2992960\\
     2996629\\
     3119256\\
     3121024\\
     3241299\\
     3327769\\
     3329136\\
     3348288\\
     3431597\\
     3441637\\
     3547874\\
     3643799\\
     3644869\\
     3655608\\
     3748392\\
     3847825\\
     4055330\\
     4058963\\
     4166365\\
     4173599\\
     4270083\\
\hline
\end{tabular}
\end{center}
\label{tab:no_detect}
\end{table}%

\section*{Figure captions}

\end{document}